\documentclass[a4paper,11pt]{article}
\usepackage{jheppub}

\usepackage{amsmath,amssymb,color,graphics,graphicx,amscd,amsfonts,epsf,ulem}
\usepackage{array}
\usepackage[utf8]{inputenc}
\usepackage{pdflscape}
\usepackage{afterpage}
\usepackage{longtable}
\usepackage{empheq}

\newcommand*\widefbox[1]{\fbox{\hspace{2em}#1\hspace{2em}}}

\newcommand{\Tr}{\ensuremath{\text{Tr}}}
\newcommand{\Ncfg}{\ensuremath{N_\text{cfg}}}

\newcommand{\tcorr}{\ensuremath{\tau_\text{corr}}}
\newcommand{\inverse}{\ensuremath{^{-1}}}
\newcommand{\order}[1]{\ensuremath{\mathcal{O}\left(    #1    \right)}}

\newcommand{\Appref}[1]{Appendix~\ref{sec:#1}}
\newcommand{\secref}[1]{Sec.~\ref{sec:#1}}
\newcommand{\Secref}[1]{Section~\ref{sec:#1}}
\newcommand{\tabref}[1]{Tab.~\ref{tab:#1}}
\newcommand{\Tabref}[1]{Table~\ref{tab:#1}}
\newcommand{\figref}[1]{Fig.~\ref{fig:#1}}
\newcommand{\Figref}[1]{Figure~\ref{fig:#1}}
\renewcommand{\eqref}[1]{(\ref{eq:#1})}
\newcommand{\Eqref}[1]{Equation~\ref{eq:#1}}
\newcommand{\Ref}[1]{Ref.~\cite{#1}}

\allowdisplaybreaks[4]

    \title{\boldmath Precision lattice test of the gauge/gravity duality at large-$N$}

    \collaboration{Monte Carlo String/M-Theory Collaboration \\ MCSMC}

    \author[a]{Evan Berkowitz}              \emailAdd{berkowitz2@llnl.gov}
    \author[a]{Enrico Rinaldi}              \emailAdd{rinaldi2@llnl.gov}
    \author[b,c,d]{Masanori Hanada}         \emailAdd{hanada@yukawa.kyoto-u.ac.jp}
    \author[e,f]{Goro Ishiki}  \emailAdd{ishiki@het.ph.tsukuba.ac.jp}
    \author[g,h]{Shinji Shimasaki}   \emailAdd{shinji.shimasaki@keio.jp}
    \author[a]{Pavlos Vranas}               \emailAdd{vranas2@llnl.gov}

    \affiliation[a]{Nuclear and Chemical Sciences Division\\ Lawrence Livermore National Laboratory, Livermore CA 94550, USA}
    \affiliation[b]{Stanford Institute for Theoretical Physics\\ Stanford University, Stanford, CA 94305, USA}
    \affiliation[c]{Yukawa Institute for Theoretical Physics\\ Kyoto University, Kitashirakawa Oiwakecho, Sakyo-ku, Kyoto 606-8502, Japan}
    \affiliation[d]{The Hakubi Center for Advanced Research\\ Kyoto University, Yoshida Ushinomiyacho, Sakyo-ku, Kyoto 606-8501, Japan}
    \affiliation[e]{Center for Integrated Research in Fundamental Science and Engineering (CiRfSE) \\ University of Tsukuba, Tsukuba, Ibaraki 305-8571, Japan}
    \affiliation[f]{Graduate School of Pure and Applied Sciences\\ University of Tsukuba, Tsukuba, Ibaraki 305-8571, Japan}
    \affiliation[g]{Research and Education Center for Natural Sciences\\ Keio University, Yokohama, Kanagawa 223-8521, Japan}
    \affiliation[h]{KEK Theory Center, High Energy Accelerator Research Organization\\ Tsukuba 305-0801, Japan}

    \preprint{
    \begin{flushright}
    LLNL-JRNL-694385, UTHEP-689, YITP-16-73
    \end{flushright}
    }

\abstract{
We pioneer a systematic, large-scale lattice simulation of D0-brane quantum mechanics. 
The large-$N$ and continuum limits of the gauge theory are taken for the first time at various temperatures $0.4 \leq T \leq 1.0$.
As a way to directly test the gauge/gravity duality conjecture we compute the internal energy of the black hole directly from the gauge theory and reproduce the coefficient of the supergravity result $E/N^2=7.41T^{14/5}$. 
This is the first confirmation of the supergravity prediction for the internal energy of a black hole at finite temperature coming directly from the dual gauge theory.
We also constrain stringy corrections to the internal energy.
}

\begin{document}
    \maketitle
    \flushbottom
    
\section{Introduction}\label{sec:intro}

The gauge/gravity duality conjecture claims that superstring theories and certain supersymmetric gauge theories are equivalent~\cite{Maldacena:1997re,Gubser:1998bc,Witten:1998qj}.
This duality implies that gauge theories provide us with a non-perturbative formulation of superstring theories, which will be essential in understanding the nature of quantum gravity.
However, this duality between gauge theories and gravity is still a conjecture.
With the aim to establish a non-perturbative formulation of superstring theories based on the duality relation, we must vigorously try to falsify the duality.

Gauge/gravity duality can be intuitively understood as a relation between two different descriptions of a system with some D-branes in a string theory.
One description of D-branes is given by the low energy effective theory of open strings, where the D-branes are described by a supersymmetric Yang-Mills theory defined on the world-volume of the D-branes.
On the other hand, D-branes can also be thought of as solitonic objects in theories of closed strings, which couple to gravity in the bulk.
In this picture, the D-branes are described as a source of gravity.
This leads to another description of the D-branes in terms of the bulk gravitational theory. 

Though the equivalence between these two descriptions is naturally expected from the physical viewpoint, no rigorous proof has been given so far.
A major obstacle is the fact that, in the duality, the perturbative semi-classical regime of superstring theory is mapped to the non-perturbative regime of the gauge theory, which is very hard to deal with in an analytical way.
In order to study the duality, one needs a method of analyzing supersymmetric gauge theories in the strong coupling regime.

Numerical simulations of gauge theories, based on lattice discretization, for example, are a powerful tool to study such a regime.
By using a discretized lattice theory, one has a robust framework to work with in order to extract information about non-perturbative physics.
This is what makes it possible to test the gauge/gravity duality from first principles.

For the duality based on D0-branes a lot of positive evidence has been obtained through numerical simulations of a supersymmetric gauge theory known as D0-brane quantum mechanics.
In this case, the gravity dual geometry is given by the black 0-brane solution in type IIA supergravity (SUGRA)~\cite{Itzhaki:1998dd}.
At finite temperature, the black 0-brane is characterized by thermodynamic quantities such as entropy and internal energy.
In particular, at large-$N$ and low temperature, where the SUGRA 
approximation becomes valid, the internal energy is given by 
\begin{align}\label{eq:energySUGRA}
  E= 7.41 N^2T^{14/5},
\end{align}
where, $E$ and $T$ are dimensionless internal energy and temperature normalized by appropriate powers of the 't Hooft coupling of D0-brane quantum mechanics. 

In this paper, we test the duality for D0-branes by performing a systematic, large-scale lattice study of D0-brane quantum mechanics.
In particular, we take both the continuum limit, by sending the lattice spacing to zero, and the large-$N$ limit for the first time.
This makes possible precise comparison with the result \eqref{energySUGRA} in the SUGRA approximation.
We calculate the internal energy of D0-brane quantum mechanics and confirm that the internal energy of the black 0-brane \eqref{energySUGRA} is reproduced from the D0-brane quantum mechanics --- our value is $E=(7.4\pm0.5)N^2T^{14/5}$.
We also give predictions for the stringy corrections directly from the gauge theory side.

The rest of this paper is organized as follows.
In \Secref{d0QM}, we review D0-brane quantum mechanics in more details and describe the existing literature.
\Secref{latticeSetup} contains the setup of our lattice simulations and the observables used to test the gauge/gravity duality.
In \Secref{results} we discuss our lattice results and their extrapolation to the continuum and large-$N$ limits, before comparing them to the SUGRA expectations in \Secref{sugraResults}.

\section{D0-brane Quantum Mechanics}\label{sec:d0QM}

We consider D0-brane quantum mechanics~\cite{Banks:1996vh},  which is the low energy effective theory of open strings ending on $N$ D0-branes in 10-dimensional flat space. 
The Lagrangian in the Euclidean signature is 
\begin{equation}\label{eq:bfss}
\mathcal{L}
=
\frac{1}{g_{YM}^2}\Tr \Bigg\{
    \frac{1}{2}(D_t X_M)^2 
-   \frac{1}{4}[X_M,X_{M'}]^2 
+   i\bar{\psi}^\alpha D_t\psi^\beta
+   \bar{\psi}^\alpha\gamma^M_{\alpha\beta}[X_M,\psi^\beta] 
\Bigg\}.
\end{equation}
Here, $X_M$ $(M=1,2,\cdots,9)$ and $\psi_\alpha$ $(\alpha=1,2,\cdots,16)$ are $N\times N$ bosonic and fermionic Hermitian matrices, the covariant derivative $D_t$ is given by $D_t=\partial_t +i[A_t,\ \cdot\ ]$ where $A_t$ is the $U(N)$ gauge field, and $\gamma^M_{\alpha\beta}$ are the left-handed part of the gamma matrices in (9+1)-dimensions, which are $16\times 16$ matrices.
This action can be obtained by dimensionally reducing the $\mathcal{N}=1$ 10D super Yang--Mills or $\mathcal{N}=4$ 4D SYM to $(0+1)$-dimension.

Historically, this model was also obtained by applying the matrix regularization to the theory of a single supermembrane in 11-dimensional flat space in the light-cone frame~\cite{deWit:1988ig}.
From this perspective, it was conjectured that the model in \Eqref{bfss} describes second quantized M-theory on 11-dimensional flat space~\cite{Banks:1996vh}.
The coupling constant $g_{YM}$ and the matrix size $N$ are related to parameters of the M-theory as $g_{YM}^2N \sim R^3 $ and $N \sim p^+ R$, where $R$ is the radius of the M-circle and $p^+$ is momentum along the light-cone direction.
In order to realize the decompactified limit $R\rightarrow \infty$ with $p^+$ fixed, one needs to take a very strong coupling limit of the matrix model.

On the other hand, in this paper, we mainly consider the 't Hooft limit of the model, where $\lambda=g_{YM}^2N$ is fixed and $N \rightarrow \infty$.
Therefore we focus on the gauge/gravity duality to type IIA superstring theory~\cite{Itzhaki:1998dd}.
The coupling constant $\lambda$ has mass dimension $3$ and sets the scale of the theory.
In the following we fix $\lambda=1$ without loss of generality, because it amounts to a rescaling of the fields.

\begin{figure}[ht]
  \centering
    \includegraphics[width=.5\textwidth]{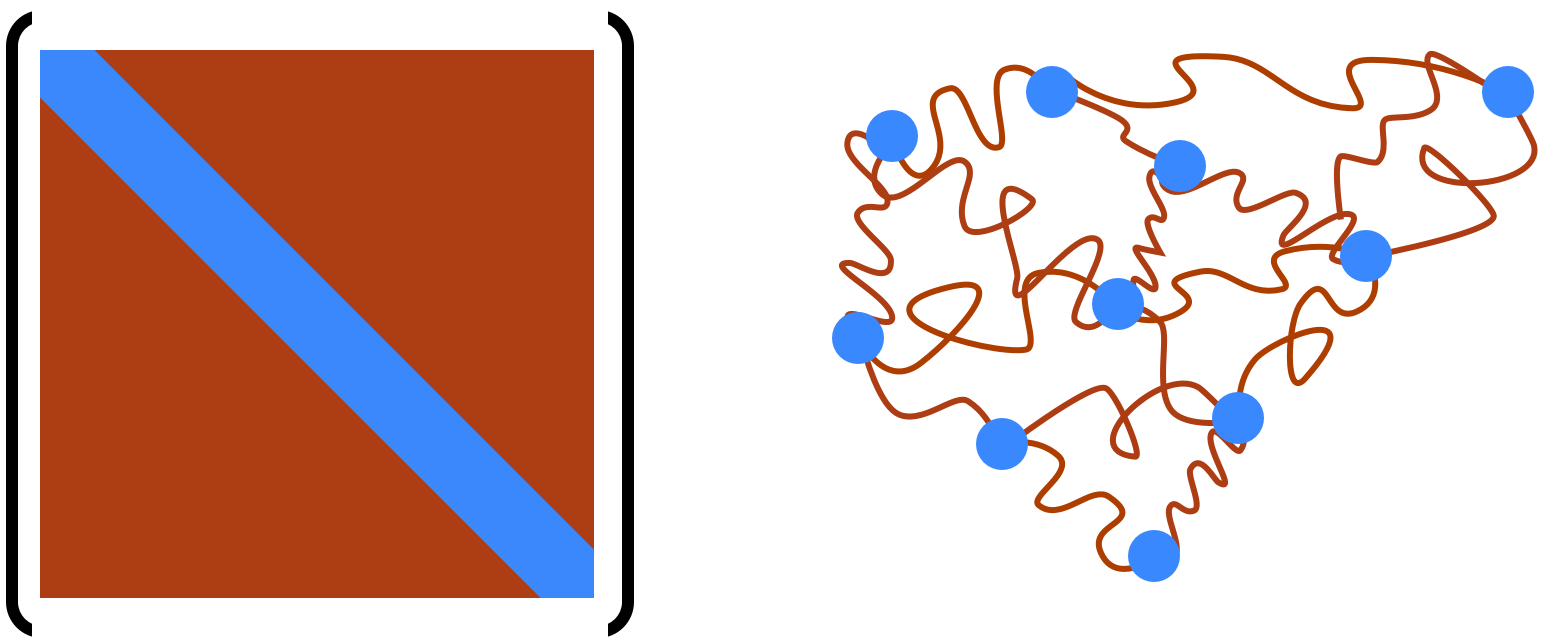}
  \caption{An intuitive interpretation of the matrices $X_M$. 
The diagonal elements correspond to positions of D0-branes and the off-diagonal elements correspond to the open strings connecting them.
This figure is taken from \Ref{Berkowitz:2016muc}.}
  \label{fig:D0-bunch}
\end{figure}

Intuitively, the off-diagonal elements of the matrices are open strings that connect the D0-branes whose locations are given by the diagonal elements~\cite{Witten:1995im}, as sketched in \figref{D0-bunch}.
Black 0-branes are states where all the D0-branes form a single bound bunch, which corresponds to generic non-commuting matrices. 
Strictly speaking, such bound state and a black 0-brane in SUGRA can be equivalent only at large-$N$ and in the strong coupling limit (low temperature\footnote{Low temperature means that the temperature $T$ is much smaller than the typical energy scale, $\lambda^{1/3}$. Hence this implies a strong coupling $\lambda^{-1/3}T\ll 1$.}).
However the bound state at generic $N$ and temperature is connected smoothly to the black 0-brane at large-$N$ and strong coupling.
Hence it can be regarded as the stringy generalization of the black hole. 
When there is no risk of confusion, we call such bound state simply as the black 0-brane or black hole.

D0-brane quantum mechanics was first investigated with Monte Carlo methods in \Ref{Anagnostopoulos:2007fw}. 
(Earlier numerical work with the same motivation can be found in \Ref{Kabat:2000zv}.) 
Previously, there have been attempts to study the internal energy~\cite{Anagnostopoulos:2007fw,Hanada:2008ez,Kadoh:2015mka,Catterall:2008yz,Filev:2015hia}, 
the supersymmetric Polyakov loop~\cite{Hanada:2008gy} and two-point correlation functions~\cite{Hanada:2009ne,Hanada:2011fq}. 

However, the existing literature claiming to provide strong evidence supporting the gauge/gravity duality could be invalidated, because the numerical results obtained so far were not extrapolated to the continuum limit $L\to\infty$ and the $N\to\infty$ limit.
This lack of controlled extrapolations would obstruct a meaningful test of the conjecture.
Moreover, the existing results did not have enough accuracy to confirm the supergravity prediction of the internal energy, $E/N^2=7.41T^{14/5}$.  
In order to achieve high precision, it is of paramount importance to correctly estimate the discretization errors and corrections due to finite $N$. 
We accomplish this for the first time in our study.

\section{Lattice Setup}\label{sec:latticeSetup}

In order to study the thermodynamic properties the D0-brane quantum mechanics non-perturbatively, 
we discretize the theory in a 0+1 dimensional Euclidean spacetime.
We then use the discretized action to calculate the theory's partition function by importance sampling field configurations via the rational hybrid Monte Carlo algorithm.
By measuring observables on this ensemble of configurations, we get an estimate for the observable's expectation value with an associated statistical uncertainty.
Finally, by measuring on ensembles with different lattice spacings, we can extrapolate to the continuum limit, removing the lattice regulator, and get a fully non-perturbative result.
As we will show, achieving a reliable continuum extrapolation requires a careful study.

\subsection{Discretized Action and Simulations}\label{sec:discretizedAction}

Consider D0-brane quantum mechanics \eqref{bfss} on a Euclidean circle with circumference $\beta$.
With antiperiodic boundary conditions for the fermions and periodic boundary conditions for the bosons, $\beta$ is identified with the inverse temperature $1/T$.

This model consists of nine $N\times N$ bosonic hermitian matrices $X_M$ ($M=1,2,\cdots,9$), sixteen 
fermionic matrices $\psi_\alpha$ ($\alpha=1,2,\cdots,16$) and the gauge field $A_t$. 
Both $X_M$ and $\psi_\alpha$ are in the adjoint representation of $U(N)$ gauge group, and the covariant derivative $D_t$ acts on them as 
$D_tX_M = \partial_t X_M +i[A_t,X_M]$ and $D_t\psi_\alpha = \partial_t\psi_\alpha +i[A_t,\psi_\alpha]$. 
The 't Hooft coupling $\lambda=g_{YM}^2N$ has a dimension of $({\rm mass})^3$, and can be set to 1 by rescaling time $t$ and the fields.
In other words, all dimensionful quantities can be made dimensionless by multiplying appropriate powers of $\lambda$.
As mentioned before, we choose $\lambda=1$.
The action is given by
\begin{equation}
S_{BFSS}=S_b+S_f, 
\end{equation}
where $S_b$ the bosonic part and $S_f$ the fermionic part are given by
\begin{align}
S_b &= N\int_0^\beta dt\ \Tr \left\{ \frac{1}{2}(D_t X_M)^2        - \frac{1}{4}[X_M,X_N]^2       \right\},   \\
S_f &= N\int_0^\beta dt\ \Tr \left\{ i\bar{\psi}\gamma^{10}D_t\psi - \bar{\psi}\gamma^M[X_M,\psi] \right\}.   
\end{align}
while $\gamma^M$ ($M=1,\cdots,10$) represent the $16\times 16$ left-handed part of the 10D gamma matrices $\Gamma^{M}$.
Formally, this model is obtained by dimensionally reducing the ten-dimensional ${\cal N}=1$ super Yang-Mills theory to one dimension. 
The index $\alpha$ of the fermionic matrices $\psi_\alpha$ corresponds to the spinor index in ten dimensions, and $\psi_\alpha$ is Majorana-Weyl in the ten-dimensional sense.

For numerical efficiency, we adopt the static diagonal gauge \cite{Hanada:2007ti}, 
\begin{equation}\label{eq:static-gauge}
A_t=\frac{1}{\beta}\cdot{\rm diag}(\alpha_1,\cdots,\alpha_N),
\qquad
-\pi<\alpha_i\le\pi. 
\end{equation} 
Associated with this gauge fixing, we add to the action the corresponding Faddeev-Popov term 
\begin{equation}
S_{F.P.}
=
-
\sum_{i<j}^{N}2\log\left|\sin\left(\frac{\alpha_i-\alpha_j}{2}\right)\right|.
\label{eq:Faddeev-Popov}
\end{equation}

We regularize the theory by discretizing the Euclidean time direction over $L$ lattice sites. Our lattice action is
\begin{align}
S_b
    &=
        \frac{N}{2a}\sum_{t,M}\Tr \left\{\left(D_+X_M(t)\right)^2\right\}
        -
        \frac{Na}{4}\sum_{t,M,N}\Tr \left\{ [X_M(t),X_N(t)]^2 \right\},
\\
S_f
    &= 
        \sum_{t}\Tr \left\{iN \bar{\psi}(t)
            \left(
            \begin{array}{cc}
            0 & D_+\\
            D_- & 0
            \end{array}
            \right)
            \psi(t)
        -
        aN\sum_{t,M}\bar{\psi}(t)\gamma^M[X_M(t),\psi(t)] \right\},  
\\
S_{F.P.}
    &=
        -
        \sum_{i<j}^{N}2\log\left|\sin\left(\frac{\alpha_i-\alpha_j}{2}\right)\right|, 
\end{align}
where the gauge links $U=\exp(iaA_t)$ with  $-\pi\le \alpha_i<\pi$.
The covariant derivative $D_{\pm}$ can be discretized in different ways which in turn will have different discretization errors.
A first discretization that we call  ``unimproved''  defines $D_{\pm}$ as follows:
\begin{align}\label{eq:Dunimproved}
  D_+f(t) &=  Uf(t+a)U^\dagger-f(t),  \nonumber\\
  D_-f(t) &=  f(t)-U^\dagger f(t-a)U ,
\end{align}
where $f(t)$ can be a bosonic or a fermionic field defined at site $t$ and the gauge link $U$ is $t$-independent due to our gauge fixing choice \eqref{static-gauge}.
This discretized derivative is related to the continuum one $D_t$ by $D_\pm f(t)= aD_t f(t) +\order{a^2}$.
The discretization of $D_{\pm}$ that we will use in our main results has smaller discretization effects, $\order{a^3}$, and we call it ``improved'' to reflect this feature.
The exact lattice definition is
\begin{align}\label{eq:Dimproved}
  D_+f(t) &=   -\frac{1}{2}U^2 f(t+ 2a)U^{\dagger2} + 2U f(t+ a)U^\dagger -\frac{3}{2} f(t),\nonumber\\
  D_-f(t) &=   +\frac{1}{2}U^{\dagger2} f(t- 2a)U^{2} - 2U^\dagger f(t- a)U +\frac{3}{2} f(t).
\end{align} 

We calculate with the unimproved and improved lattice actions with the RHMC algorithm, tuning the integration step and trajectory length to attain an acceptance rate of order $80\%$.
We take advantage of MPI parallelization, where each MPI process takes care $l$ lattice sites and $n\times n$ sub-blocks of matrices.
The number of total processes for a lattice of size $L$ and matrices of size $N \times N$ is $(L/l)\cdot(N/n)^2$.
Typically we take $n=l=4$ and, for example, the number of processes is $8^3=512$ for $N=L=32$.
This setup is very advantageous on large parallel machines and allows us to simulate very large values of $N$ and $L$ by scaling our code to greater numbers of MPI processes.
The simulation code is publicly available and well documented \cite{simulation_code}.

An important remark for numerical simulations of the D0-brane quantum mechanics is that the system has flat directions, $[X_M,X_{M'}]=0$. 
At large $N$, the flat directions are lifted dynamically, around the black hole phase. 
However, at finite $N$, the black hole is metastable, and the D0-branes (the eigenvalues of the matrices) can be emitted and propagate to infinity. 
This phenomenon produces an instability in the Monte Carlo evolution which become more and more severe at smaller $N$ and at lower temperatures.
In order to obtain meaningful statistical results from simulations, it is of crucial importance to control these flat directions and correctly single out the phase under consideration \cite{Anagnostopoulos:2007fw,Hanada:2013rga,Hanada:2010qg}.
If this control is missing, wrong answers might be obtained, as happened countlessly many times in the early literature on lattice supersymmetry.
In this study, we overcome the instability by taking $N$ sufficiently large that our observables do not show signs of eigenvalue instability over long Monte Carlo histories.

\subsection{Observables}\label{sec:observable}

On each configuration we measure different observables.
The most crucial for this work is the internal energy $E/N^2$,
\begin{equation}\label{eq:internalenergy}
    E/N^2 = \frac{3}{2N^2\beta}\left(    9(N^2L-1) - 2 \langle S_b\rangle    \right). 
\end{equation}

We also measure the absolute value of the Polyakov loop,
\begin{equation}\label{eq:polyloopabs}
    |P| = \left|\frac{1}{N}\sum_{j=1}^N e^{i\alpha_j}\right|
\end{equation}
where $\alpha_j$ belong to the gauge-fixed link variables, the average size of the eigenvalue bunch (0-brane) $R^2$, 
\begin{equation}\label{eq:size0brane}
   R^2 \equiv \frac{1}{NL} \sum_{M,t} \Tr\left\{ X_M^2\right\} \qquad\longrightarrow\qquad  \frac{1}{N\beta}\int dt \Tr X_M^2
\end{equation}
and the potential term $F^2$ (analogous to the square of the field strength),
\begin{equation}\label{eq:potentialterm}
   F^2 = -\frac{1}{NL}\sum_{M,M',t} \Tr\left\{\left[X_M,X_{M'}\right]^2\right\}
    \qquad\longrightarrow\qquad  -\frac{1}{N\beta}\int dt \Tr [X_M,X_{M'}]^2. 
\end{equation}

\subsection{Phase Quenching}\label{sec:phasequench}

One potential issue in simulating this theory is the infamous sign problem --- the Pfaffian that results from integrating out the fermions can have an oscillating phase, undermining the probabilistic interpretation of the Euclidean action in the path integral.
In our calculation, we follow the usual practice~\cite{Anagnostopoulos:2007fw,Hanada:2008ez,Catterall:2008yz,Filev:2015hia,Kadoh:2015mka} of simply taking the absolute value of the Pfaffian, quenching the phase.

Several studies have found that the phase of the Pfaffian remains close to zero in the temperature region we consider, and the most recent one is \Ref{Filev:2015hia}.
This means that the sign problem is mild and quenching the phase does not distort the results.
Previous results were obtained for relatively small values of $N$ and of the cutoff, but in the same temperature regime we study here.

In principle, the effect of the phase can be taken into account by phase reweighting, 
\begin{eqnarray}\label{eq:phase_quench}
\langle {\cal O}\rangle_{\rm F}
=
\frac{\langle {\cal O}\cdot e^{i\theta}\rangle_{\rm PQ}}{\langle e^{i\theta}\rangle_{\rm PQ}}, 
\end{eqnarray}
where $\langle\ \cdot\ \rangle_{\rm F}$ and $\langle\ \cdot\ \rangle_{\rm PQ}$ represent the expectation values with the full and phase-quenched theories, and $e^{i\theta}={\rm Pfaffian}/|{\rm Pfaffian}|$. 
Interestingly, even when the phase fluctuations become large, it has been observed that the phase quenching does not affect the expectation values of various observables.

\begin{figure}[tb]
  \centering
    \includegraphics[width=.65\textwidth]{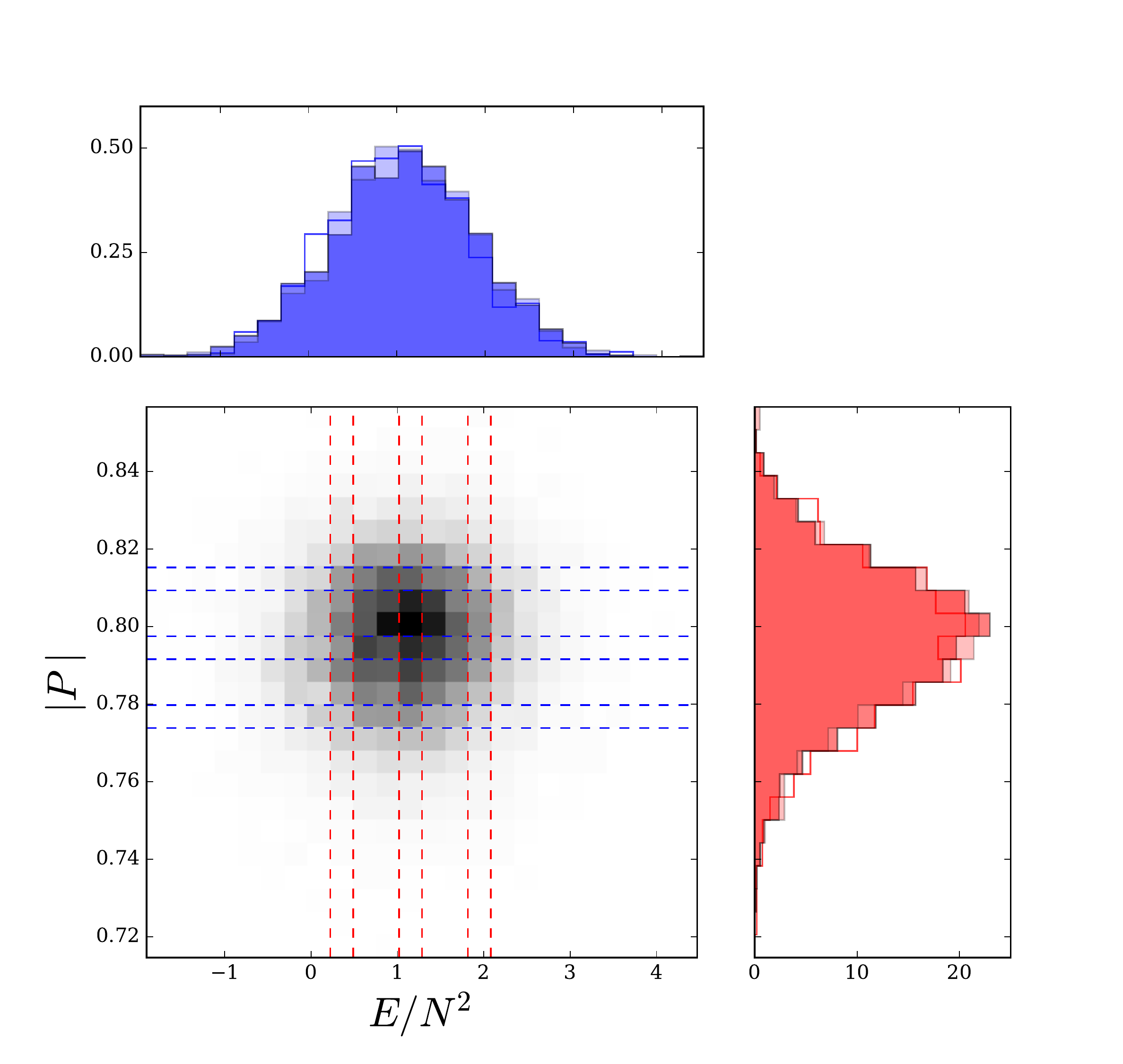}
  \caption{ The correlation between $E/N^2$ and $|P|$ at $N=16$, $L=16$ and $T=0.5$ is shown as 2D histogram where a darker color corresponds to a higher count.
    The blue and red histograms (three per panel) represent the normalized distribution of $E/N^2$ and $|P|$, respectively, within the slices on the two-dimensional plot bounded by dashed lines.
  The histograms coming from different slices are almost indistinguishable.}\label{fig:E-P_correlation}
\end{figure}

A possible mechanism is suggested in \Ref{Hanada:2011fq}.
Let $\rho(x)$ be the distribution of the observable ${\cal O}$ in the phase-quenched simulation, and let $w_x$ be the average of $e^{i\theta}$ when the value of ${\cal O}$ is fixed to $x$.
Then
\begin{align}
  \label{eq:phase-quenching}
  \langle {\cal O}\rangle_{\rm PQ} &
                                     =\int dx\ x \rho(x)
\\
  \langle {\cal O}\cdot e^{i\theta}\rangle_{\rm PQ} &
                                                      =\int dx\ x \rho(x) w_x
\\
  \langle e^{i\theta}\rangle_{\rm PQ} &
                                        =\int dx\ \rho(x) w_x
\ .
\end{align}
Typically $\rho(x)$ peaks around the average value, $x=\langle {\cal O}\rangle_{\rm PQ}$. 
If $w_x$ is constant around this peak, then $\langle {\cal O}\cdot e^{i\theta}\rangle_{\rm PQ}\simeq \langle {\cal O}\rangle_{\rm PQ}\cdot \langle e^{i\theta}\rangle_{\rm PQ}$, and then \eqref{phase_quench} becomes $\langle {\cal O}\rangle_{\rm F}\simeq \langle {\cal O}\rangle_{\rm PQ}$. 

Because the calculation of the Pfaffian is very costly, it is difficult to test this scenario directly at large values of $N$.
However, it is possible to indirectly infer the magnitude of the phase fluctuations and their impact on the other observables. 
In fact, the Polyakov loop has a strong correlation with the phase factor---the phase disappears when $|P|=1$ (up to discretization effects) and the phase fluctuations become larger as $|P|$ decreases. 
In \figref{E-P_correlation}, we show the correlation between $E/N^2$ and $|P|$ at $N=16$, $L=16$ and $T=0.5$.
The blue and red histograms represent the distribution of $E/N^2$ and $|P|$, respectively, with the other quantity restricted within small bins highlighted in the two-dimensional plot. 
The areas of the histograms are normalized. 
The $E$-independence of the distribution of $|P|$, at least away from the tails, strongly suggests the $E$-independence of the distribution of the phase, which justifies the phase quenching via the scenario explained above. 
A more detailed study of the distribution of $|P|$ for various values of the energy, near and away from its average, is reported in \Appref{correlations}.

An explicit calculation of the Pfaffian phase is worthwhile, but we leave it for a future study.
In the rest of the paper we assume that the phase-quenched approximation does not influence the internal energy results of our simulations.

\section{Results}\label{sec:results}

In this section we discuss the statistical needs of our analysis, continuum extrapolations at fixed $N$ (comparing with other calculations when available) and simultaneous continuum and large-$N$ extrapolations.
In the end we report a continuum large-$N$ data set that we will use in \Secref{sugraResults} for a direct comparison to supergravity predictions.
We also collect our measurements for each ensemble in \Appref{measurements}.

\subsection{Statistical Requirements}

\begin{figure}[htbp]
    \centering
        \includegraphics[width=\textwidth]{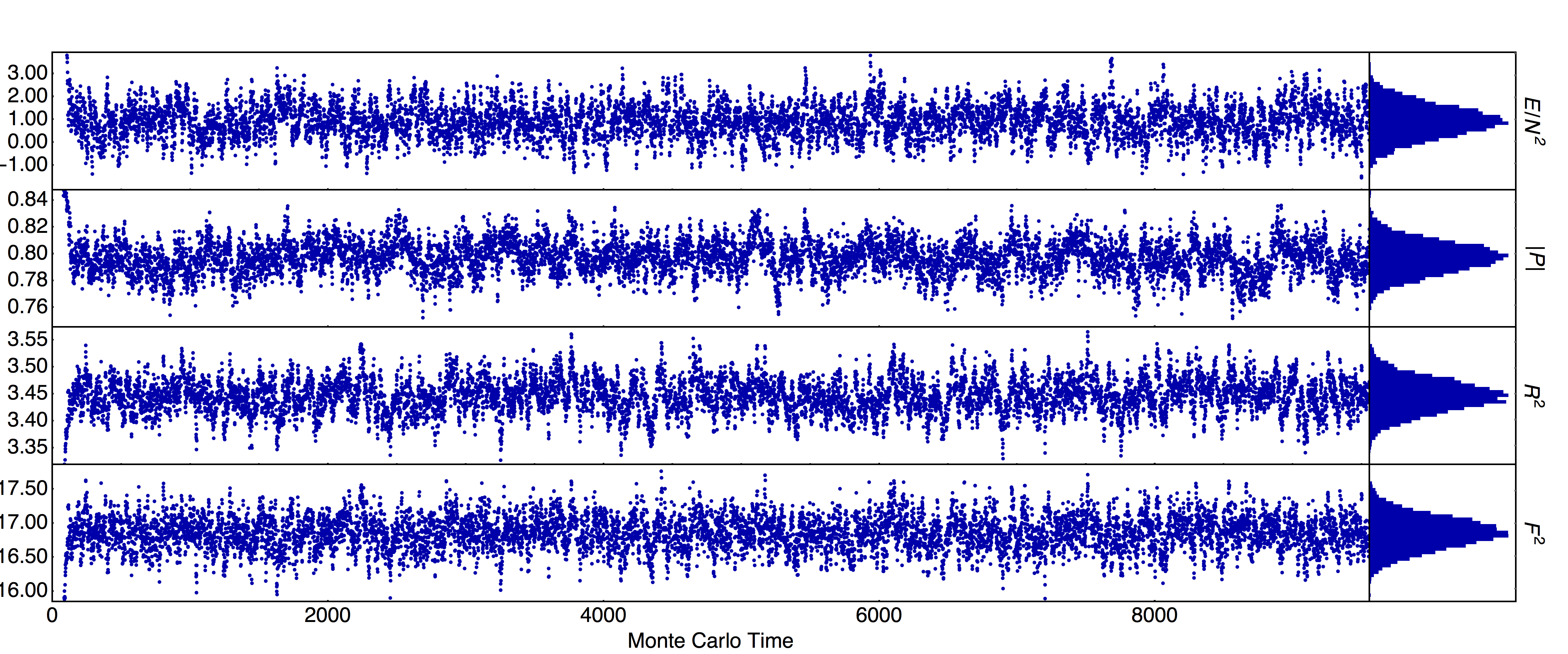}
    \caption{
    The Monte Carlo history and a corresponding histogram for the energy $E/N^2$, the Polyakov loop $|P|$, $R^2$, and $F^2$ of the  $T=0.5$ $N=24$ $L=32$ ensemble.
    For each observable, one can see fluctuations that span many Monte Carlo steps.}
    \label{fig:mc}
\end{figure}

To ensure a faithful estimation of an observable, one must ensure a large number of independent (decorrelated) Monte Carlo samples are taken.
In \figref{mc} we show an example Monte Carlo history for the ensemble with $N=24$, $L=32$, and $T=0.5$.
It is apparent that there are long-lived autocorrelations.
Therefore, to achieve many independent samples, lengthy Monte Carlo ensembles are required.

\begin{figure}[htbp]
    \centering
        \includegraphics[width=0.49\textwidth]{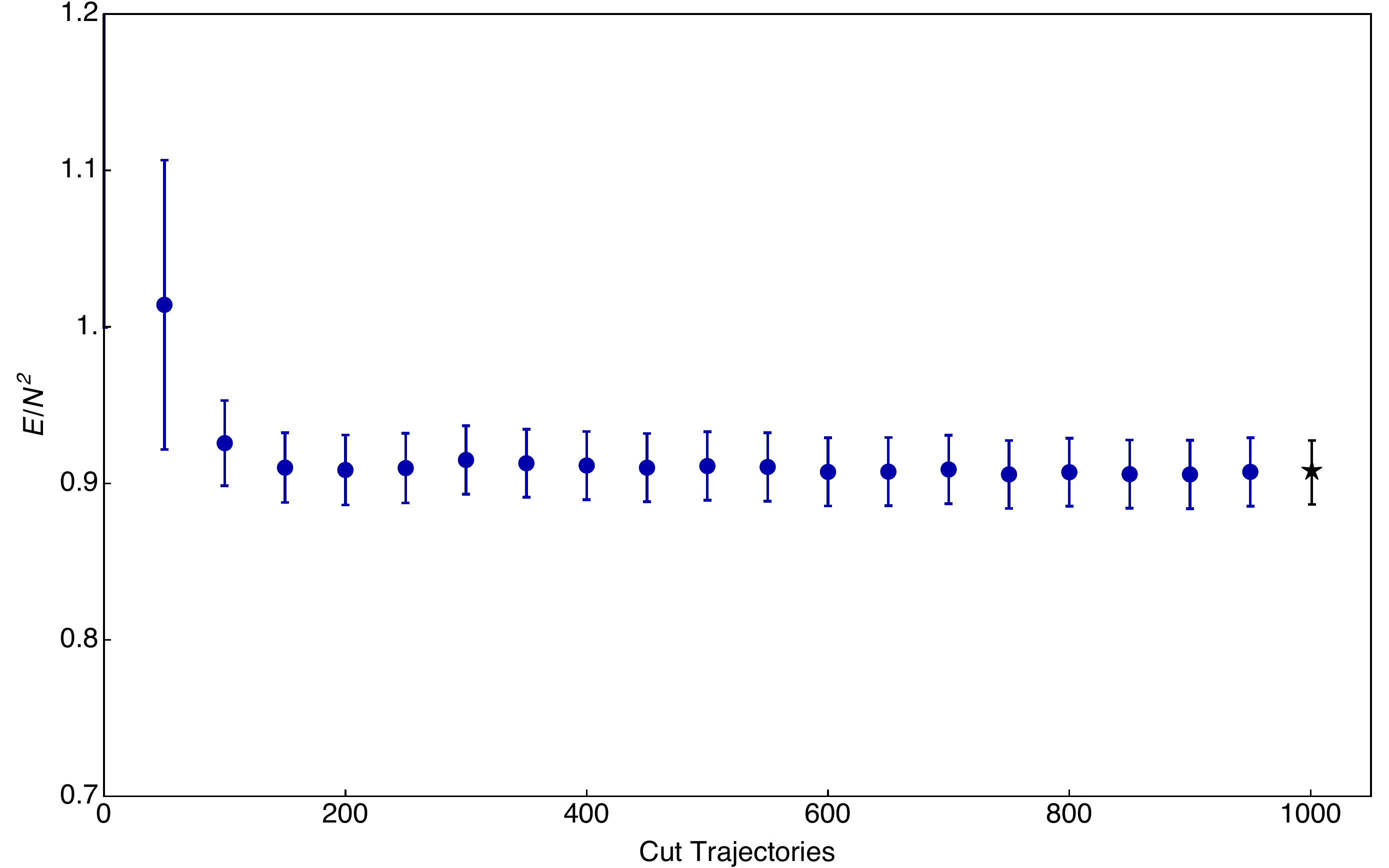}
        \includegraphics[width=0.49\textwidth]{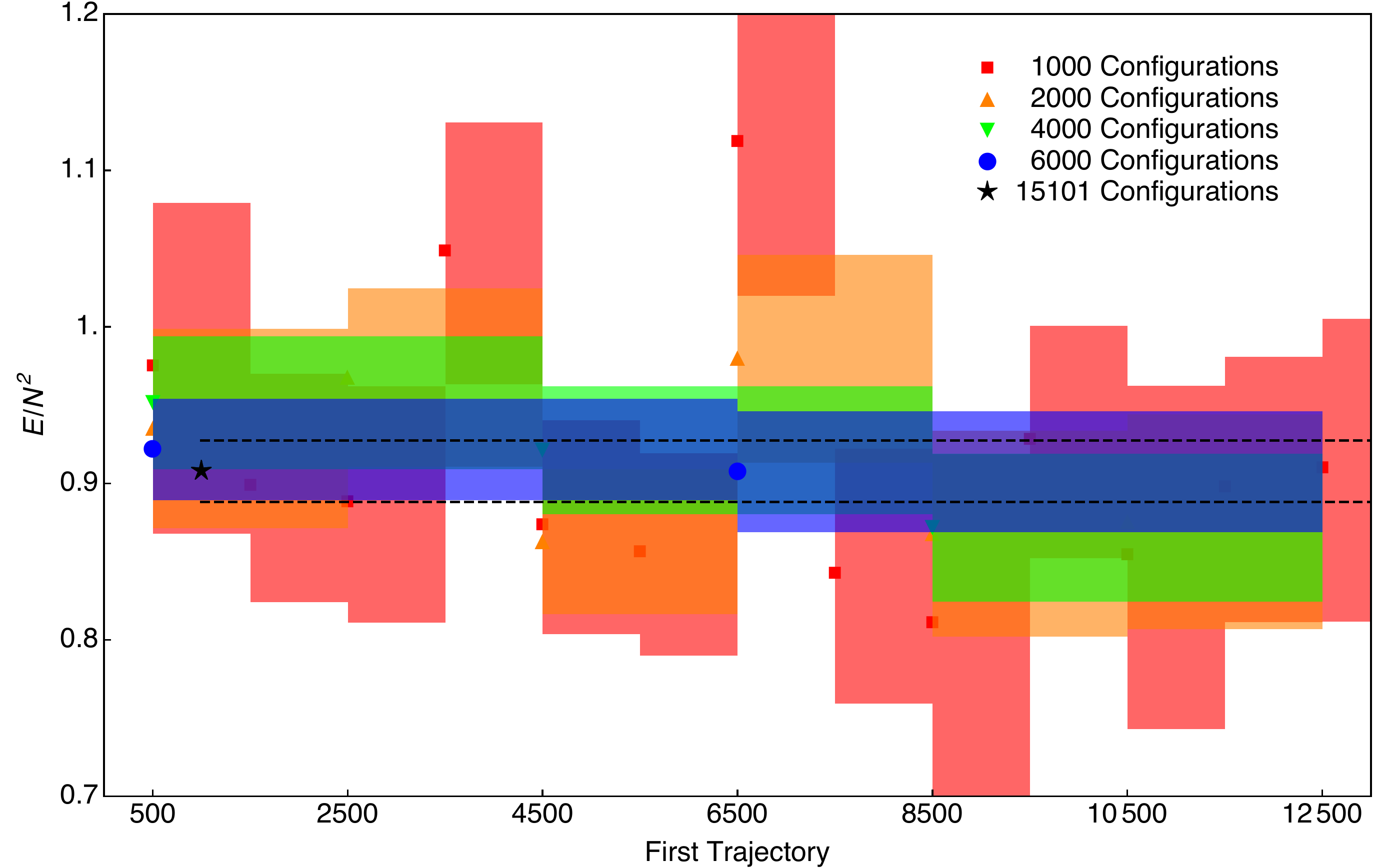}
    \caption{
    A study of the statistical stability of $E/N^2$ for the ensemble $N=16$, $L=32$, $T=0.5$.
    In the left panel we show different thermalization cuts, measuring $E/N^2$ on the rest of the configurations.
    In the right panel we show the importance of large statistical samples by measuring on consecutive disjoint sets of trajectories.
    As the statistical sample grows from 1000 configurations (red squares) to 6000 configurations (blue circles), the central values and uncertainties between sets of configurations become more and more stable and compatible.
    In both panels we perform the analysis with bins of 50 configurations.
    For comparison, we also show our final analysis and its uncertainty as a black star in both panels, with its error bar displayed as dashed lines on the right panel.
}
    \label{fig:stats}
\end{figure}

Moreover, accounting for autocorrelations is essential for an accurate estimate of the statistical uncertainty on a given measurement.
For each observable on each ensemble, we measure the autocorrelation time \tcorr\ using the Madras-Sokal algorithm and form bins of width 3\tcorr.
With those binned measurements we perform a jackknife analysis to estimate the statistical uncertainty.
We also independently test that the statistical error associated with our final average is robust by performing different analysis with smaller and larger jackknife bins and making sure that the final uncertainty does not change.

In \figref{stats} we study the statistical stability of $E/N^2$ for a low-temperature ensemble, $N=24$, $L=32$, and $T=0.5$, with bins 50 trajectories wide.
In the left panel we show the residual effects of keeping measurements from too early in the Monte Carlo history by using the whole ensemble and only adjusting the thermalization cut.  
From the compatibility with later cuts, it is clear that this ensemble has no memory of its initially chosen configuration after 500 trajectories.
On each ensemble, we discard 1000 trajectories as a thermalization cut.

In the right panel of \figref{stats} we show how many configurations are necessary for a stable estimate.
We start at trajectory 500 and take the next 1000, 2000, 4000, and 6000 trajectories and perform an independent analysis, and then slide that window to the next disjoint set of trajectories.
One can see that for this ensemble, 1000 thermalized trajectories is not enough to achieve a stable statistical estimate, indicating that there can be sizable fluctuations over Monte Carlo time that can dramatically shift the measured value.
However, 2000 trajectories seem to be enough to reliably get the eventual central value within the uncertainty.
Increasing the window size correctly washes out the effect of lengthy fluctuations and makes each successive analysis agree more reliably.
We are therefore confident that most of our statistical samples are large enough to correctly estimate the energy $E/N^2$.

Some ensembles at $T=0.4$ are not very lengthy---though all are longer than 1000 trajectories after the thermalization cut.
To compensate for this shortcoming, we inflated their statistical uncertainty by 50\% and reperformed all the following analyses.
We find very little difference between the two cases.
In what follows, we therefore use the uninflated errors.
\subsection{Continuum Extrapolation at Fixed $N$}
\begin{figure}[thb]
    \centering 
    \includegraphics[width=0.65\textwidth]{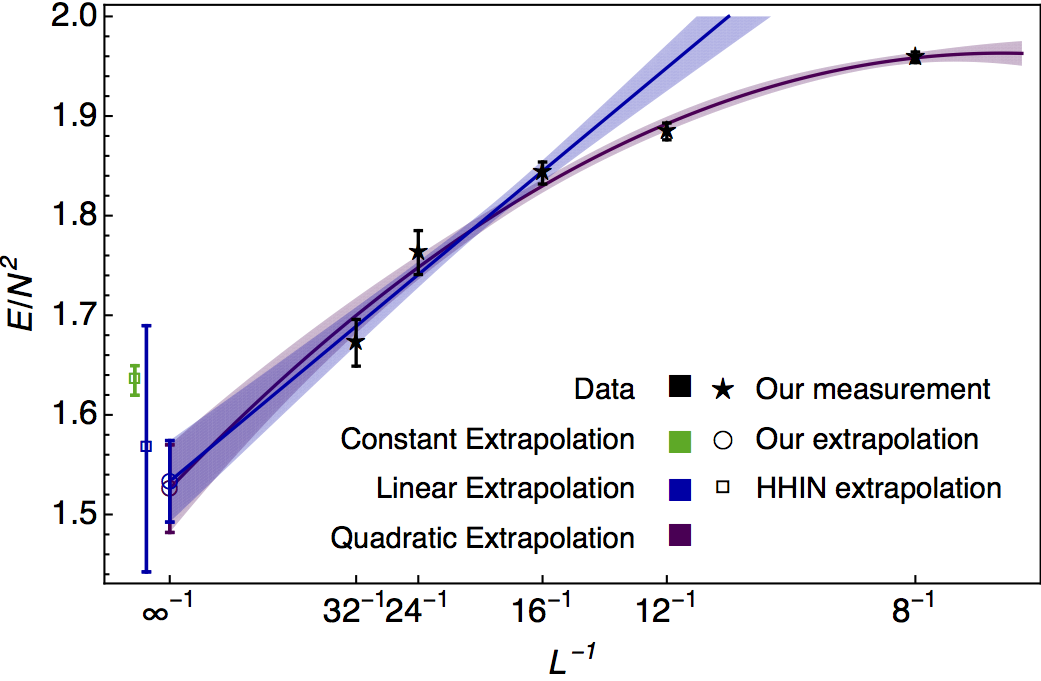}
    \caption{A continuum extrapolation for $T=0.7$ with fixed $N=16$ with the improved action.  Black stars represent our measurements, open circles our extrapolation, and open squares the extrapolation from \Ref{Hanada:2013rga}.
    Error bars and the error bands on the extrapolated curves represent 1$\sigma$ errors.
    Our linear extrapolation is only fit to the data at $L\geq16$.
    The divergence of the linear and quadratic extrapolations indicate that for this ensemble, linear extrapolations of lattice data that include data taken at $L\lesssim 16$ will be systematically biased.
    }
    \label{fig:fixed_N}
\end{figure}

To study the continuum theory, one must measure at a variety of lattice discretizations and extrapolate to the continuum.
In this section, we discuss continuum extrapolation at fixed $N$ using our unimproved and improved actions (cfr. equations \eqref{Dunimproved} and \eqref{Dimproved}).

As the lattice spacing $L\inverse$ gets smaller, one expects an expansion around $L\inverse=0$ to get better.
So, at fixed $N$ the energy should follow
\begin{equation}\label{eq:extrapolation_fixed_N}
    \frac{E}{N^2} = e_0 + \frac{e_1}{L} + \frac{e_2}{L^2} + \order{L^{-3}}
\end{equation}
where $e_0$ is the continuum-extrapolated value and the other $e_i$ characterize the lattice artifacts.
Based on the na\"ive scaling of the action with the lattice spacing, we expect results with the unimproved action to have larger discretization effects and we check this explicitly in the following for the first time.

In \Figref{fixed_N} we show a fixed-$N$ continuum extrapolation for $T=0.7$ $N=16$ so that we can directly compare to the continuum extrapolations of \Ref{Hanada:2013rga}.
One immediately sees that the region where only the leading $L\inverse$ corrections matter is $L\gtrsim16$---with smaller $L$ the subleading correction is not negligible, so linear fits to lattice data from such small $L$ will be systematically biased towards larger $E/N^2$.
We have checked this rule of thumb for all $T$ and $N$, and find broad consistency with this observation, which means Refs.~\cite{Hanada:2013rga,Filev:2015hia} may suffer from premature extrapolation.

\begin{figure}[thb]
    \centering 
    \includegraphics[width=0.65\textwidth]{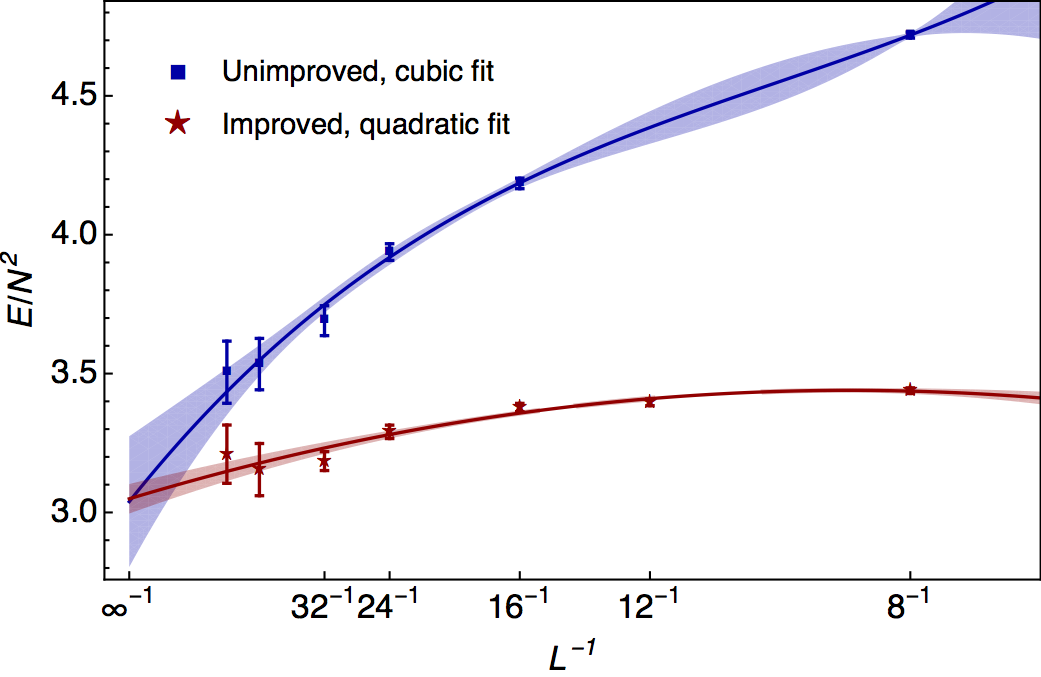}
    \caption{A comparison between taking the continuum limit for the unimproved and improved actions for $T=1.0$, $N=16$.
    Error bars and the error bands on the extrapolated curves represent 1$\sigma$ errors.
    }
    \label{fig:improvement}
\end{figure}

Knowing that to successfully fit down to $L=8$ with the improved action requires a quadratic fit, we expect additional lattice artifacts to contaminate $L=8$ with the unimproved action, suggesting an additional term is needed to fit the unimproved action to incorporate that point into the continuum limit.
Indeed, fitting a quadratic to that point pushes the fit upwards, while fitting a cubic gives perfect agreement with the improved continuum limit.
Using the improved action allows us to extrapolate to the continuum in a more controlled manner, because a successful extrapolation requires fitting fewer parameters.

\subsection{Simultaneous Large $N$ and Continuum Extrapolation}
\label{sec:simultaneous}
\begin{figure}[thb]
    \centering
        \includegraphics[width=\textwidth]{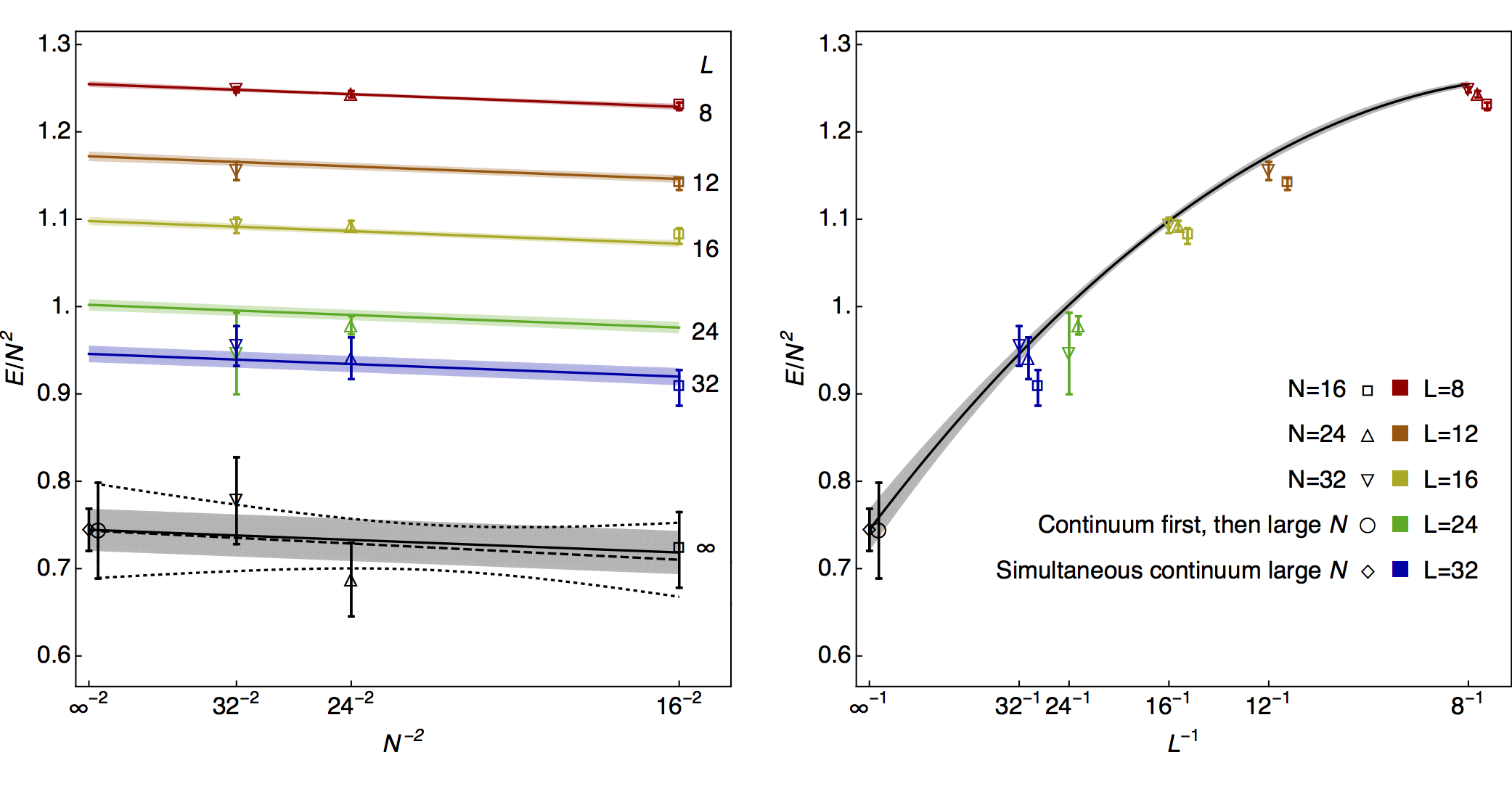}
    \caption{
    A simultaneous continuum- and large-$N$ extrapolation for $T=0.5$ with the improved action.
    All the data points are fit to a single 2D surface given by \eqref{2D_fit_form}.
    In the right panel, we show all the data points (slightly offset for visual clarity) and the (black) $N=\infty$ slice of the fitted surface.
    In the left panel, we show all the data points and the (black) continuum extrapolations at each $N$ together with their subsequent large-$N$ limit as the dashed line with uncertainties given by the dotted band.
    We also show the fixed-$L$ slices of the 2D fit, as well as the (black) $L=\infty$ slice.
    In both panels, the black circle represents the result of taking the large-$N$ limit of the continuum extrapolations at each $N$, while the best-fit simultaneous continuum- and large-$N$ limit $e_{00}$ is shown as a black diamond.
    Error bars, the error band on the extrapolated surface, and the dotted error band represent 1$\sigma$ uncertainties.
    }
    \label{fig:extrap}
\end{figure}

In order to test the gauge/gravity duality precisely, it is important to take the large-$N$ limit.
However, taking the continuum limit at large-$N$ becomes costly even with a quadratic fit, because at small $N$ the physical instability may ruin the Monte Carlo history, while numerical cost grows with $L$ and $N$.

Large-$N$ corrections appear in powers of $N^{-2}$, at each fixed $L$, because the 't Hooft counting holds even for the discretized theory.
Thus, at a fixed temperature we expect $E/N^2$ to be described by a series like the following 
\begin{equation}\label{eq:2D}
    \frac{E}{N^2} = \sum_{i,j\geq0} \frac{e_{ij}}{N^{2i}L^j}
\end{equation}
so that $e_{i0}$ are physical, continuum-limit quantities at finite $N$, $e_{00}$ is the continuum, large-$N$ value, and all other coefficients, $e_{ij}$ with $j>0$, characterize lattice artifacts.
Importantly, by extrapolating in $1/N^2$ and $1/L$ simultaneously, we can take advantage of significantly more data points without increasing the number of fit parameters dramatically.

We can truncate \eqref{2D} in various ways and attempt to fit a finite set of $e_{ij}$.
We attempted a six-parameter fit with $i+j\leq2$ and found our data insufficient to characterize $e_{11}$ or $e_{20}$ without 100\% uncertainties, and strong correlation with the other coefficients.
We also performed five-parameter fits, omitting either $e_{11}$ and $e_{20}$ and still found the other to be very poorly constrained by our data and highly correlated with the remaining coefficients.
Thus, we settled on a four-parameter fit---next to leading order (NLO) in $N^{-2}$ and NNLO in $L\inverse$, with no mixed term
\begin{equation}\label{eq:2D_fit_form}
    \frac{E}{N^2} \approx e_{00} + \frac{e_{01}}{L} + \frac{e_{02}}{L^2} + \frac{e_{10}}{N^2}.
\end{equation}
We fit this form to all of our measurements at a given temperature, and find extremely good fit quality together with a very mild dependence on $N$ and---just as in the fixed-$N$ case---important dependence on $L$.

The strong $L$ dependence, which we observe to get stronger at low temperature, raises the possibility that \Ref{Kadoh:2015mka}, which at low temperature works only at $L=16$ and has no continuum limit, and \Ref{Hanada:2013rga}, which at $N=16$ extrapolates from the momentum cutoff $\Lambda\leq8$, may be systematically contaminated by discretization artifacts.
However, because those references use different discretized actions from that used in this work, their discretization effects may be substantially smaller than in our approach.
For example, a direct comparison at $T=0.4$ $N=32$ $L=16$ shows that \Ref{Kadoh:2015mka}'s central value is substantially closer to our continuum limit 0.40(7) than our data point at those parameters 0.835(7).

In \Figref{extrap} we show the result of the simultaneous continuum- and large-$N$ extrapolation of the measurements of the $T=0.5$ improved action measurements.
We also show three fixed-$N$ continuum extrapolations and their subsequent large-$N$ extrapolation.
For that ensemble, we fit 13 data points to the four-parameter fit in \eqref{2D_fit_form} and find a reduced chi-squared (the usual $\chi^2$ divided by DOF, the degrees of freedom in the fit) of 7.2/9 and good compatibility with the sequential extrapolation.

In \Tabref{extrap} we show the simultaneous continuum and large-$N$ extrapolation by the four-parameter fit in \eqref{2D_fit_form} of data taken with the improved action at various temperatures.
A more complete data set is provided in \Appref{extrap_full}.

\begin{table}[th]
    \begin{center}
\begin{tabular}{c>{$}r<{$}@{$\pm$}>{$}l<{$}>{$}r<{$}@{$\pm$}>{$}l<{$}rc}
 $T$ & \multicolumn{2}{c}{$e_{00}$} &  \multicolumn{2}{c}{$-e_{10}$} & $\chi^2$ & DOF    \\\hline
 0.4 &  0.38 & 0.06                 &  5.4 & 9.2                     & 1.3      &  4 \\
 0.5 &  0.74 & 0.02                 &  6.7 & 1.5                     & 7.2      &  9 \\
 0.6 &  1.15 & 0.02                 &  5.0 & 1.8                     & 8.8      &  8 \\
 0.7 &  1.54 & 0.03                 &  3.9 & 2.0                     & 8.8      &  8 \\
 0.8 &  1.99 & 0.03                 &  6.2 & 2.5                     & 15.1     &  8 \\
 0.9 &  2.57 & 0.04                 & 11.9 & 2.9                     & 3.3      &  8 \\
 1.0 &  3.11 & 0.04                 &  8.4 & 3.2                     & 8.9      & 10 \\
\end{tabular}
    \end{center}
    \caption{
The continuum energy coefficients $e_{00}$ (large-$N$) and $e_{10}$ (leading $1/N^2$ correction) for different temperatures, the $\chi^2$ of the extrapolating fit, and the degrees of freedom for that fit.
In every case $\chi^2$/DOF is between 0.3 and 1.9.
    }
    \label{tab:extrap}
\end{table}

\section{Supergravity and Black Hole Internal Energy}\label{sec:sugraResults}
To ultimately check the gauge/gravity duality, we want to compare our gauge-theory calculations with supergravity (SUGRA) and superstring calculations.
As is thoroughly reviewed in \Ref{Hanada:2013rga}\footnote{The study of $\alpha'$ and $g_s$ corrections based on string perturbation theory has a long history.
In the example of the type IIA black 0-brane under consideration, the $\alpha'$ expansion corresponds to the expansion with $\alpha'/R_{BH}^2\sim T^{3/5}$, where $R_{BH}$ is the curvature radius of the black hole geometry.
At tree level, it starts with $(\alpha')^3$~\cite{Gross:1986iv,Gross:1986mw,Grisaru:1986px,Grisaru:1986vi} and is followed by $(\alpha')^5, (\alpha')^6, \cdots$~\cite{Green:1999pv,Green:2006gt}. 
A more detailed argument including $g_s$ corrections can be found in \Ref{Hyakutake:2013vwa}, \Ref{Hyakutake:2014maa} and references therein.}
, the internal energy of the black 0-brane can be expanded with respect to $T$ and $1/N^2$ as
\begin{align}\label{eq:internal_energy}
    \frac{E}{N^2} 
    &= 
        \frac{\left(    a_0 T^{14/5} + a_1 T^{23/5} + a_2 T^{29/5} + a_3T^{32/5}+\cdots    \right)}{N^0}
    +   \frac{\left(    b_0 T^{2/5} + b_1 T^{11/5} +               \cdots    \right)}{N^2}
    +   \order{\frac{1}{N^4}}   \nonumber \\
    &=  \frac{E_{0}(T)}{N^0} + \frac{E_{1}(T)}{N^2} +   \order{\frac{1}{N^4}}
\end{align}
where $a_0$ and $b_0$ are known by exact calculations to be approximately $7.41$ and $-5.77$ respectively.
We group the coefficients at a fixed order in $N$ into the functions $E_i(T)$.
On the gauge-theory side of the duality, these functions should be reproduced by our coefficients $e_{i0}$ reported in \tabref{extrap}.

In this section we will present a variety of fits comparing our extrapolated values in \tabref{extrap} to these forms and we will summarize our findings in the next section.
At each temperature, we have access to the continuum large-$N$ behavior ($E_0$ through $e_{00}$) and the $1/N^2$ correction ($E_1$ through $e_{10}$) independently.
This allows us to fit the different orders of $N^2$ in \eqref{internal_energy} separately.
\subsection{SUGRA at Low Temperatures}\label{sec:confirmation}
\begin{figure}[htbp]
    \centering
        \includegraphics[width=0.65\textwidth]{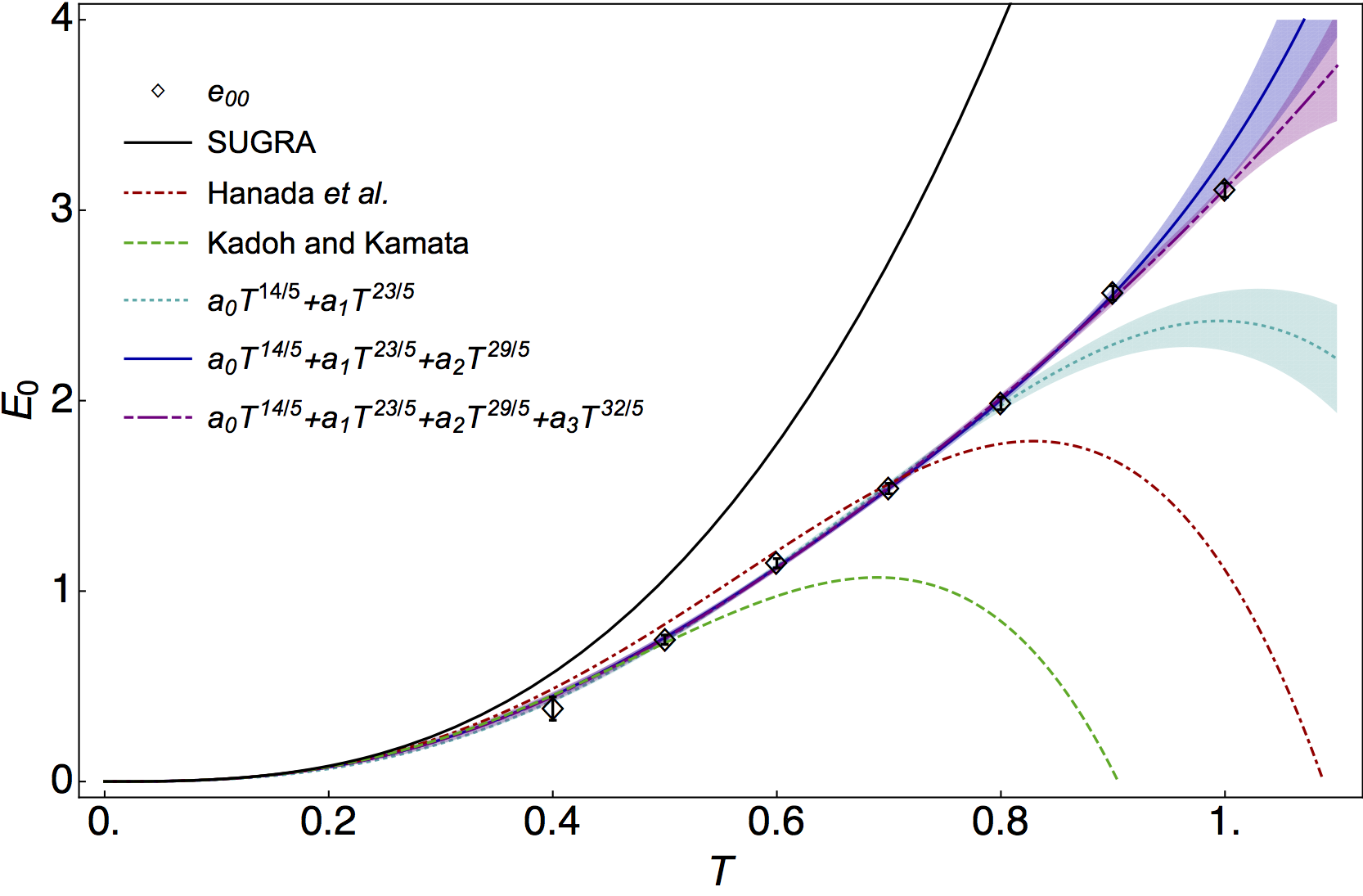}
    \caption{
    Our best fits of $E_0$ to the data points $e_{00}$ shown as black diamonds, including the first two/three/four terms as a cyan dotted line/blue solid line/purple dot-dot-dashed line with $1\sigma$ error bands.
    We also show the result from \Ref{Kadoh:2015mka} and \Ref{Hanada:2008ez} results as green dashed and red dot-dashed lines, respectively.  The SUGRA result is shown in black.}
    \label{fig:sugra}
\end{figure}

First, let us confirm that our continuum and large-$N$ data are consistent with the SUGRA prediction, in that they reproduce the SUGRA calculation of the leading coefficient $a_0=7.41$.
The agreement between D0-brane quantum mechanics and SUGRA is our main result.
Checking the value of $a_0$ against lattice simulations is a non-trivial task and is usually hindered by numerical results with large error bars or with undefined systematic errors. 

We fit the $\order{N^0}$ coefficients, including the leading-order coefficient known from supergravity.
We perform two fits of $E_{0}$ to $e_{00}$, fitting $a_0$ and $a_1$, including or excluding $a_2$.
We exclude the $T=1.0$ data point, because the assumption $T\ll1$ is certainly broken there.
In \figref{sugra} we show the best fits of $E_0(T)$, together with previous estimates of the same function and the SUGRA result.

The fit that excludes $a_2$ struggles to capture the full behavior of the data, and in the best case (fitting to $T\leq0.8$) produces $a_0=6.2\pm0.2$, substantially different from the supergravity result, and $a_1=-3.8\pm0.3$.
However, this can be understood as a systematic issue---trying to capture too much temperature dependence without including the next term of $E_0$ distorts the coefficients.
The term with the $a_2$ coefficient is as important as the one with $a_1$ and should be fitted together in the temperature region of our data.
In fact, also fitting $a_2$ relieves the tension between the two terms and produces 
\begin{empheq}[box=\widefbox]{align*}
        a_0&=7.4\pm0.5   
    &   a_1&=-9.7\pm2.2
    &   a_2&=5.6\pm1.8
    &   \chi^2/\text{DOF} &= 2.6/3.
\end{empheq}
Adding an additional term $a_3$ representing a higher order $\alpha'$-correction does not modify the above results, while the uncertainties increase dramatically.

Our value for $a_0$ is entirely consistent with the SUGRA-predicted value of $7.41$ and has a very small uncertainty $\sim 7\%$.
This agreement may be considered a \textit{bona fide} direct test of the gauge/gravity duality: if the D0-brane quantum mechanics and supergravity results differed, we could have falsified the correspondence.

\begin{figure}[htbp]
    \centering
        \includegraphics[width=0.65\textwidth]{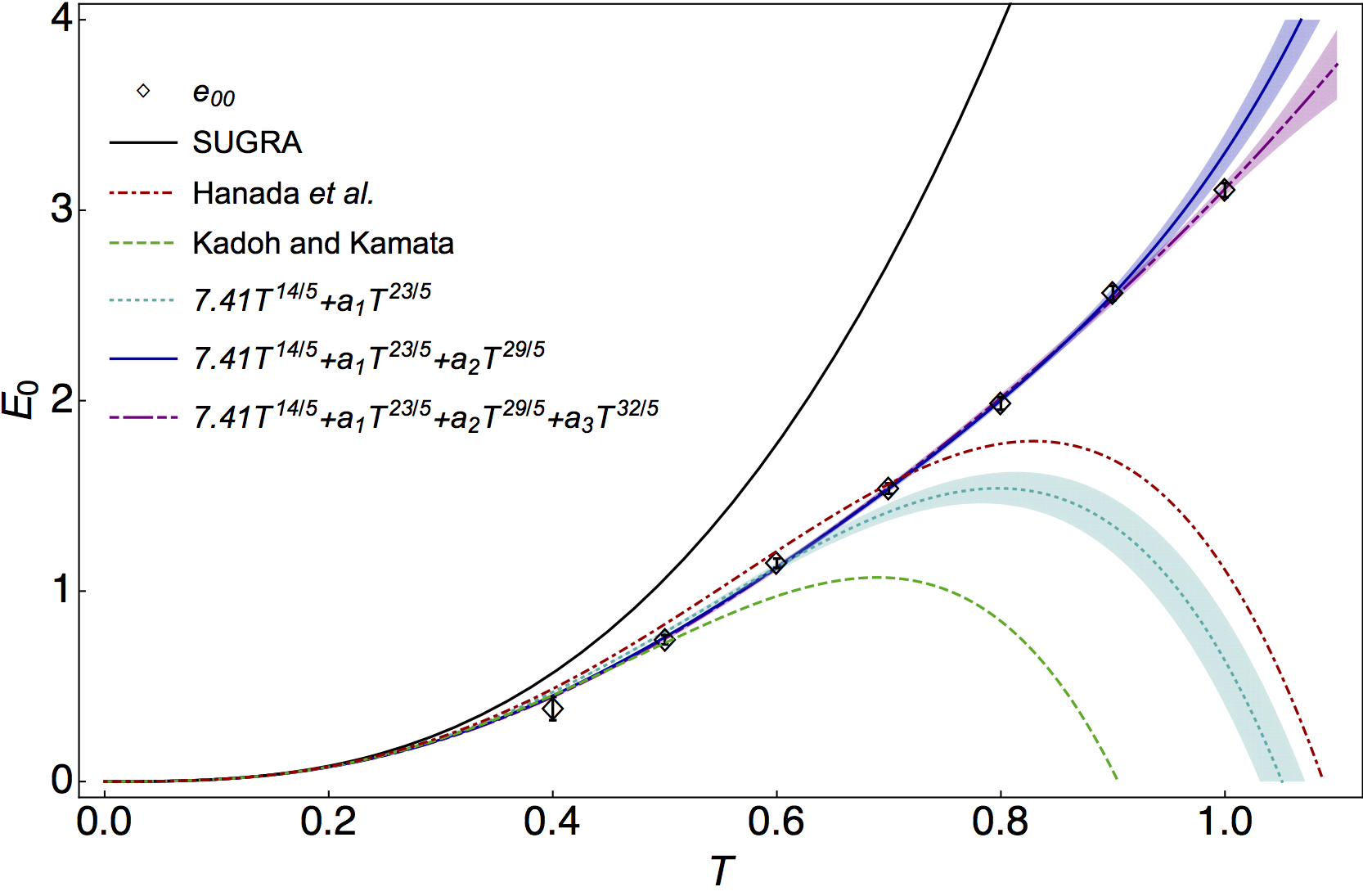}
    \caption{
    The same as in \figref{sugra}, but with $a_0$ fixed to its known SUGRA value, rather than fit.
    }
    \label{fig:subleading_T}
\end{figure}

We have also fit $a_1$ and $a_2$ while fixing the known SUGRA value $a_0=7.41$.
If we exclude $a_2$, we do not get a good fit, even if we change the fitting window.
Including $a_2$ dramatically improves the fit---we can comfortably incorporate all the data up to $T < 1$.  
The best fit is to $T\leq0.9$ and gives $a_1=-10.0\pm0.4$ and $a_2=5.8\pm0.5$, describes the data well ($\chi^2$/DOF=2.6/4), and is in very good agreement with our best fit when we did not demand the SUGRA value $a_0=7.41$---further bolstering our confidence in that result.\footnote{A fit of this form including $a_3$ reproduces a compatible value for $a_1$ but fails to reproduce $a_2$. We find that $a_2+a_3$ for this fit is compatible with the central value $a_2=5.8$ of our previous fit. This is not surprising because the corresponding powers $T^{29/5}$ and $T^{32/5}$ are very close.}

The full form of $E_0$ in \eqref{internal_energy} is actually, on the gravity side, an expansion in $\alpha'/R_{BH}^2=T^{3/5}$ where $\alpha'$ is the string coupling and $R_{BH}$ the black hole radius.
That is, generically
\begin{equation}\label{eq:power_series}
    E_0 = A_0 T^{14/5} + A_1 T^{17/5} + A_2 T^{20/5} + A_3 T^{23/5} + A_4 T^{26/5} + A_5 T^{29/5} + \cdots
\end{equation}
However, the coefficients $A_{1,2,4}$ are known to vanish based on string theory calculations.

We tried a variety of strategies to verify from our data that those coefficients do indeed vanish.
We performed a 6-parameter fit to our 7 data points, tried fixing $A_0$ to its known value, tried fixing $A_0$, $A_3$ and $A_5$ to the best-fit values of \secref{confirmation}.
In no case did we get a reliable fit, nor could we empirically confirm that these coefficients vanish.
This is unsurprising, because to distinguish the terms we need to sample temperatures where, for example, $T^{14/5}$ and $T^{17/5}$ differ notably---which is difficult in the temperature range of our data. 
Indeed, we are fortunate that those terms vanish, because it is much easier to distinguish the different nonvanishing powers (as we did at the beginning of this section) when those powers are more widely separated.
Obtaining information at smaller temperatures becomes crucial in order to determine higher order corrections more precisely.

\subsection{Subleading Temperature Dependence}\label{sec:nnlo}
\begin{figure}[htb]
    \centering
        \includegraphics[width=0.65\textwidth]{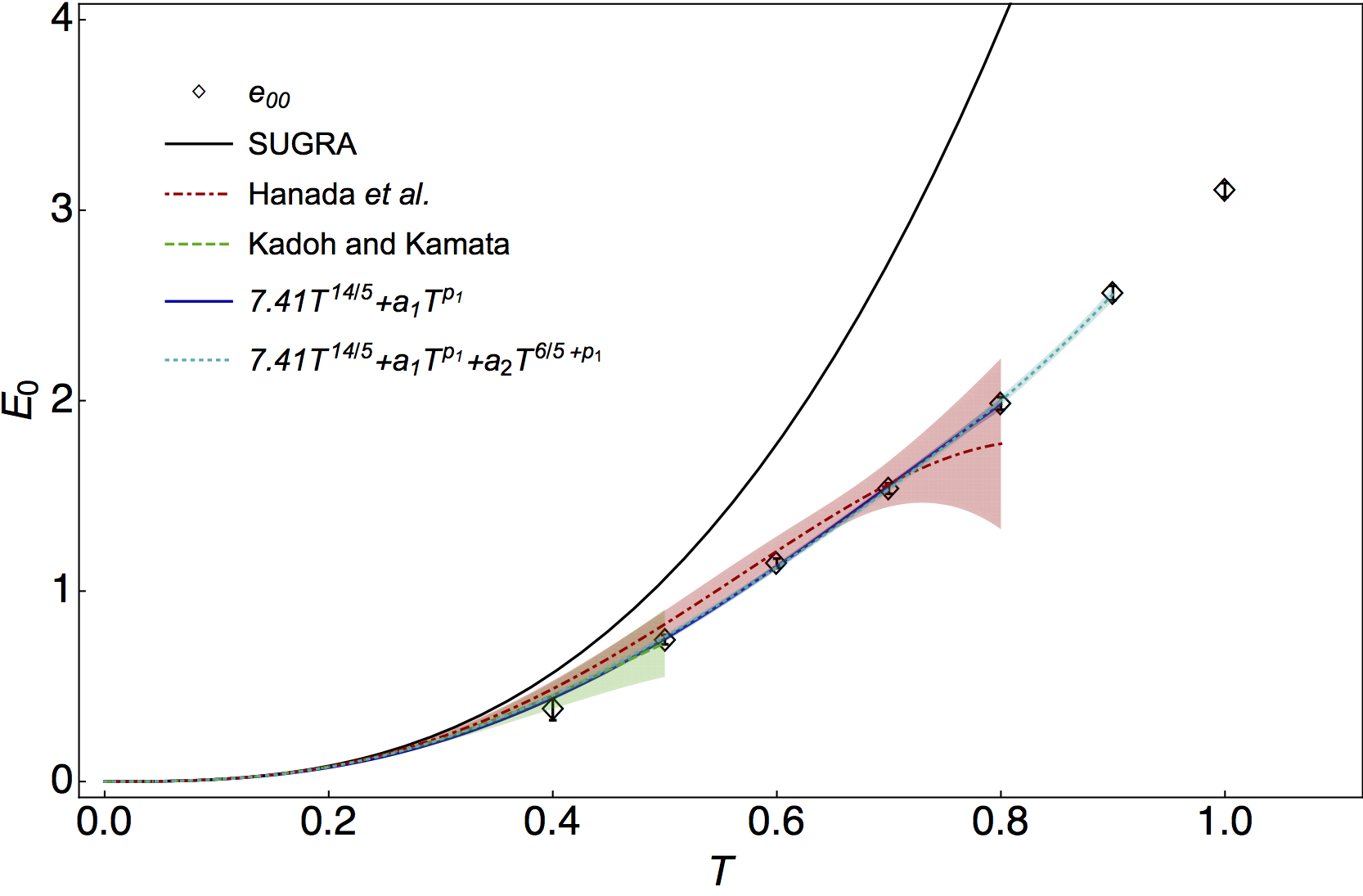}
    \caption{
    Two fits of our continuum large-$N$ values $e_{00}$ (black diamonds) for $E_0(T)$.
    The solid blue line is a fit to \eqref{nlop} over $0.4\leq T \leq 0.8$, while the dotted cyan line is a fit to \eqref{n2lop} over $0.4\leq T \leq 0.9$, with their respective (small) error bands.
    We also show the results from Hanada et al.~\cite{Hanada:2016zxj} and Kadoh and Kamata~\cite{Kadoh:2015mka} as red dot-dashed and green dashed lines, respectively, with their 1$\sigma$ uncertainty band as explained in the main text.
    The SUGRA result is shown in black.
    }
    \label{fig:nlop}
\end{figure}

Previous work has fit the form
\begin{equation}\label{eq:nlop}
    E_0(T) = 7.41 T^{14/5} + a_1 T^{p_1}
\end{equation}
where the exponent of the first correction is also unconstrained, but with a fixed leading behavior.
Our best fit to this form is for $0.4\leq T \leq0.8$, yielding $a_1=-4.7\pm0.2$ and $p_1=3.9\pm0.1$ ($\chi^2$/DOF=1.6/3).
We are unable to reproduce the known power $p_1=23/5=4.6$ from this fit, indicating that the temperatures used in the fit were too high to identify this dependence alone, or the temperature range is too wide for the data to be described by simply the next-to-leading-order power of $T$.\footnote{It is interesting to note that, if we fit our data by a single power law $E_0(T)=aT^p$, then $E_0(T) = (3.13\pm 0.03) T^{2.02\pm0.03}\approx\pi T^2$ describes our continuum large-$N$ data very well ($\chi^2$/DOF=7.7/5) in the whole temperature range.
We emphasize that this may be a coincidence.}
In fact, trying to incorporate the next nonzero $\alpha'$ correction by fitting
\begin{equation}\label{eq:n2lop}
    E_0(T) = 7.41 T^{14/5} + a_1 T^{p_1} + a_2 T^{p_1+6/5}
\end{equation} 
produces 
\begin{empheq}[box=\widefbox]{align*}
        p_1&=4.6\pm0.3
    &   a_1&=-10.2\pm2.4
    &   a_2&=6.2\pm2.6
    &   \chi^2/\text{DOF} &= 2.6/3.
\end{empheq}
These values for $a_1$ and $a_2$ match very well with the results of the previous section, where all the powers were fixed, and $p_1$ matches the predicted value exactly.
This fit takes advantage of the knowledge that on the gravity side the energy can be characterized by a power series in $T^{3/5}$ as explained in \eqref{power_series} and that some of the coefficients vanish.

To avoid incorporating knowledge from the string theory side, we would prefer to fit the different powers independently rather than requiring them to differ by 6/5.
However, executing such a fit is extremely tricky without imposing the qualitative requirement that the two exponents differ nontrivially.
This requires a more sophisticated analysis.

In a calculation with the momentum cutoff regularization~\cite{Hanada:2007ti} at $N\le 17$, \Ref{Hanada:2008ez} obtained $a_1=-5.55(7)$ and $p_1=4.58(3)$ by using data points at $0.5\le T\le 0.7$, without an explicit estimate of the discretization errors the effect of $N$ finite.
More recently, the continuum limit at $N=16$ has been studied in \Ref{Hanada:2016zxj} and we use those continuum, fixed-$N$ results to perform an additional fit to \eqref{nlop} which is reported as ``Hanada et al.'' in \figref{nlop}.
The resulting parameters are consistent with the ones at fixed cutoff~\cite{Hanada:2008ez}, and the function $E_0(T)$ overlap with all our data points due to the large uncertainty of the fit.
However, with our lattice data extrapolated to the continuum and large-$N$ limit we have demonstrated that the next-to-leading order temperature dependence can not be singled out with accuracy, without accounting for the next $\alpha'$ correction.

In another lattice study described in \Ref{Kadoh:2015mka}, the authors obtain $a_1=-9 \pm 2$ and $p_1= 4.74\pm 0.35$ from data at $0.375 \leq T \leq 0.475$, again without a continuum limit or an extrapolation to large $N$.
In \figref{nlop} we show how their results compare with our data points and the other fits in the literature.
In the same range of temperatures we only have one continuum value for $e_{00}$ which hinders our ability to reproduce their result from \eqref{nlop}.\footnote{In \Figref{nlop} the uncertainty band for the fit to the data points in \Ref{Hanada:2016zxj} is obtained by actually performing the fit to the published data, while we rely on a private communication with the authors of \Ref{Kadoh:2015mka} for the uncertainty band in that case.}

\subsection{$\mathcal{O}\left(    N^{-2}    \right)$ Corrections}\label{sec:N2_physics}
\begin{figure}[hbt]
    \centering
        \includegraphics[width=0.45\textwidth]{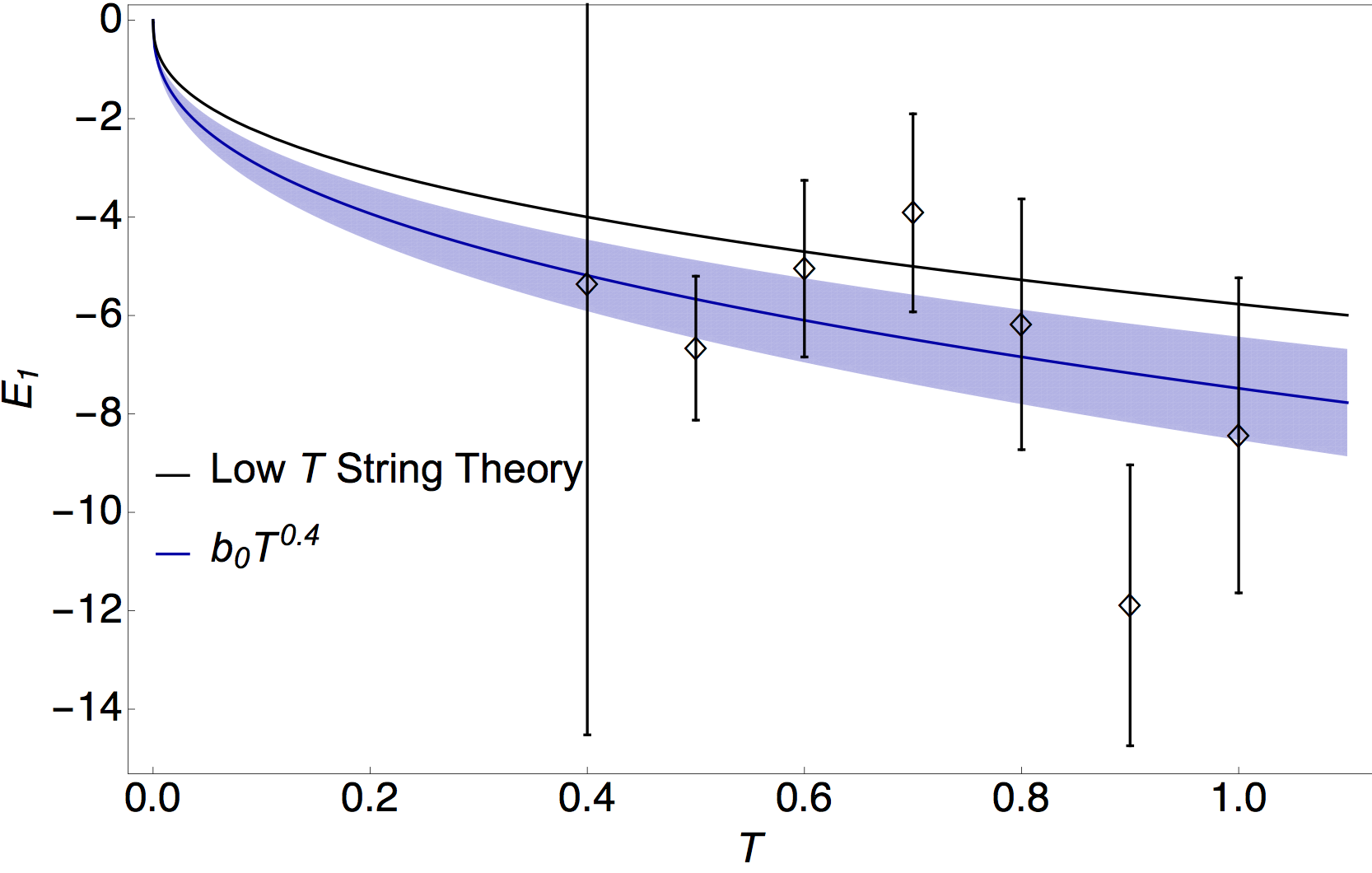} 
        \includegraphics[width=0.45\textwidth]{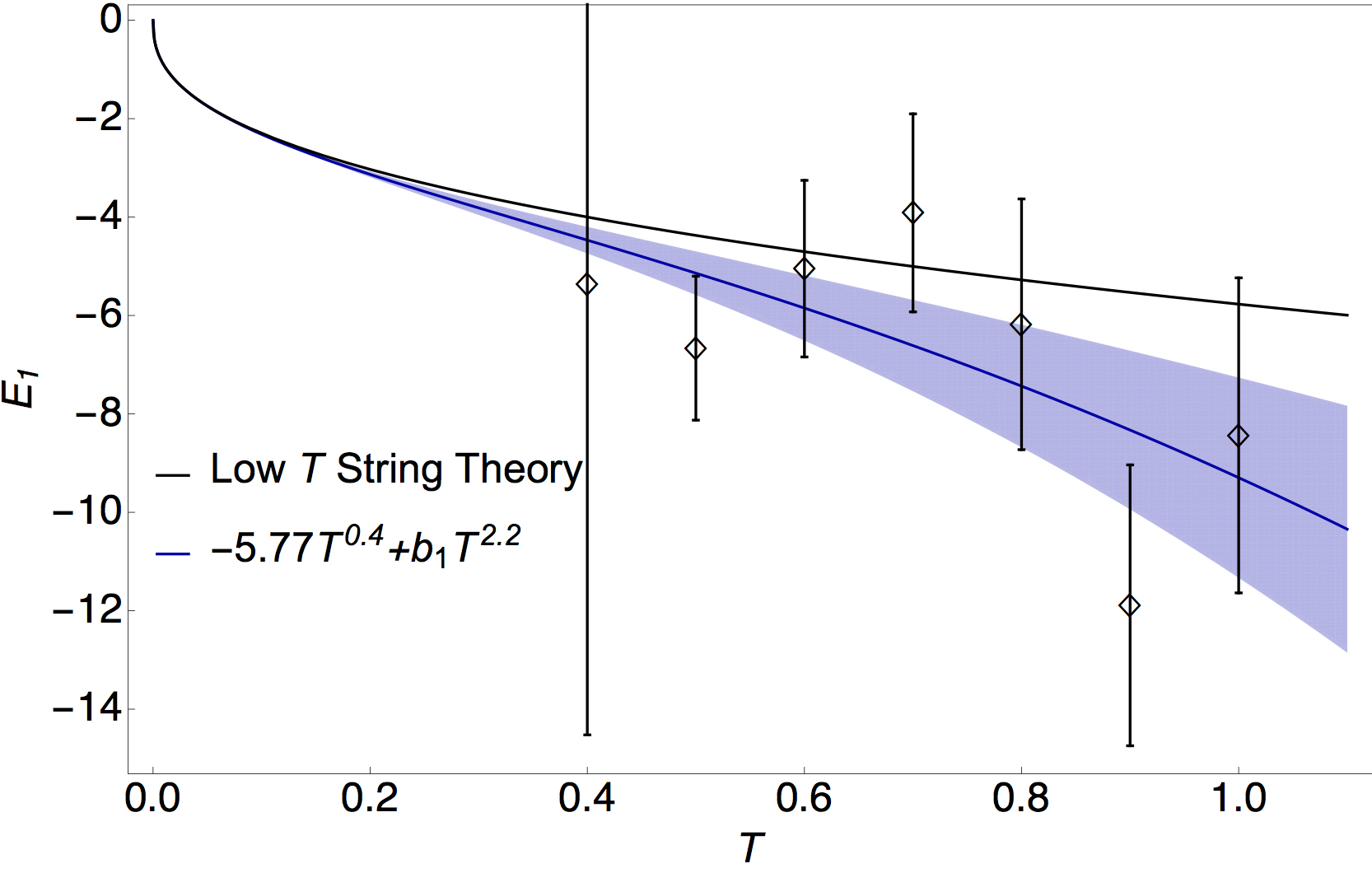}  \\
        \includegraphics[width=0.45\textwidth]{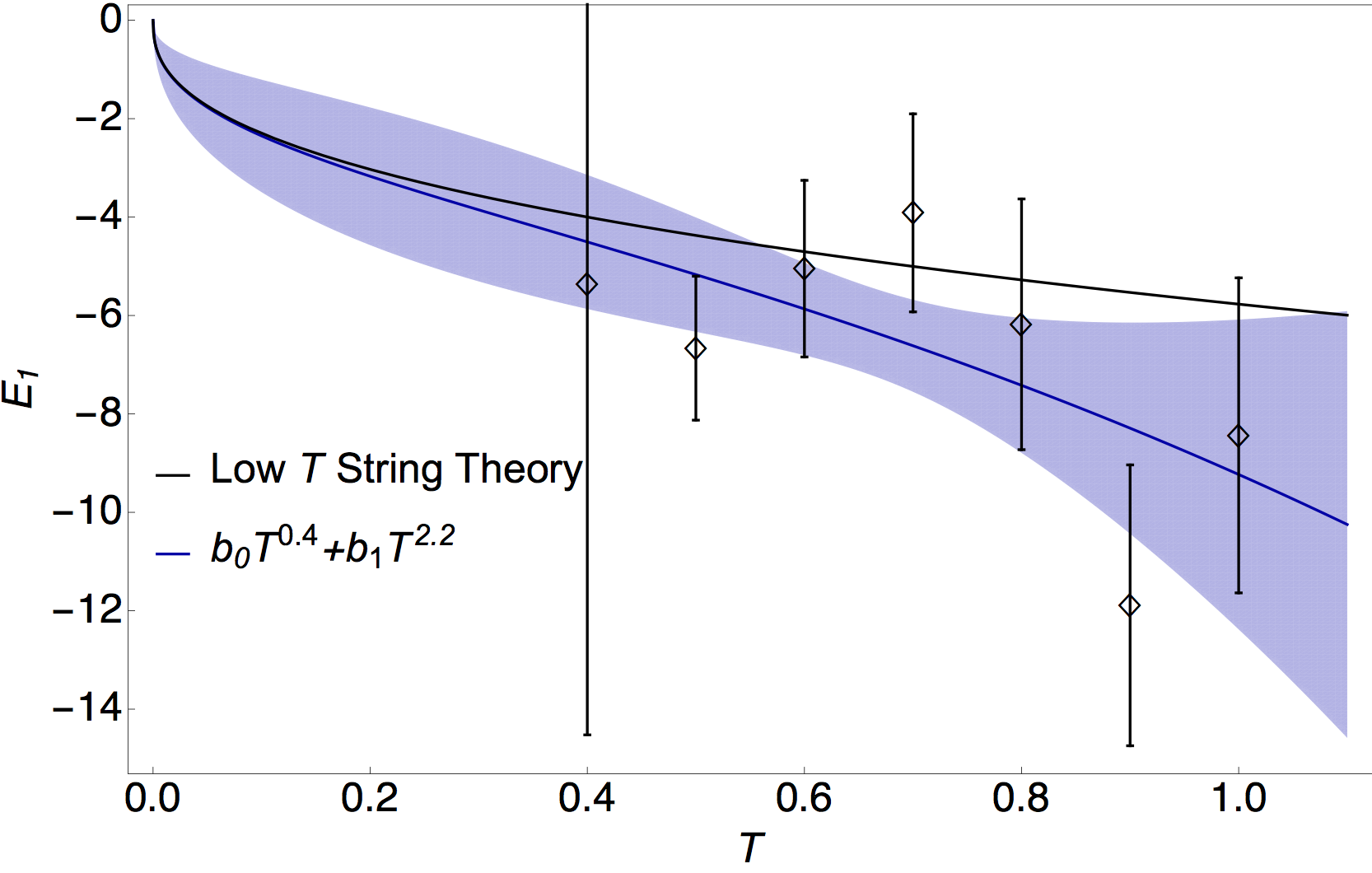}
        \includegraphics[width=0.45\textwidth]{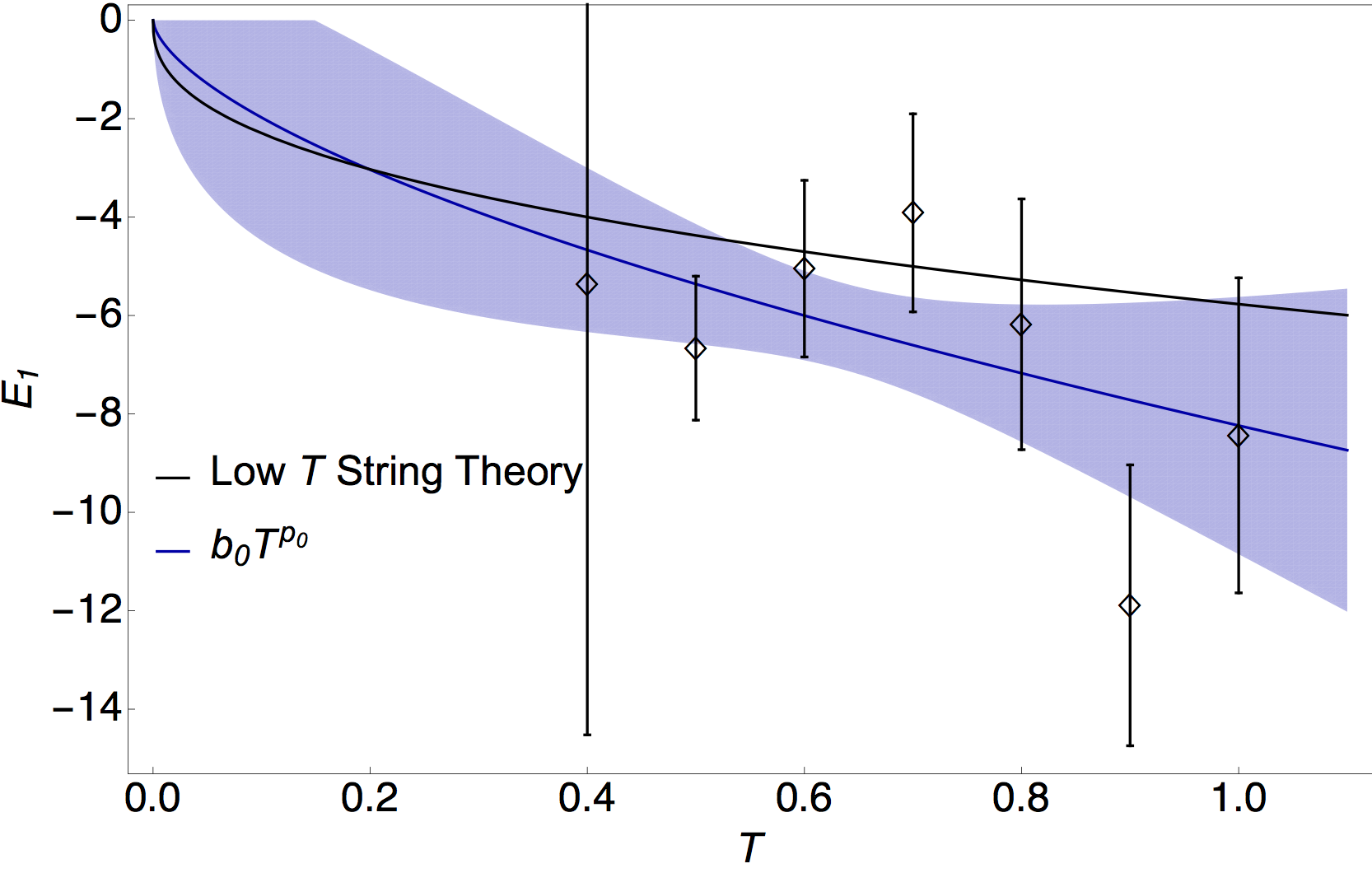}
    \caption{
    Four fits of $E_1$ to our values for $e_{10}$.
    In each panel, we show our measurements as black diamonds with 1$\sigma$ error bars, the fit as the solid blue curve with a 1$\sigma$ error band, and the known low-temperature behavior $b_0=-5.77$ as the black curve.
    In the top left  panel we fit just $b_0$, with $b_1=0$.
    In the top right panel we fix $b_0=-5.77$ and fit $b_1$.
    In the bottom left  panel we fit both $b_0$ and $b_1$.
    In the bottom right panel we fit $b_0T^p$.
    All fits have $0.9<\chi^2$/DOF$<1.3$, and cannot be meaningfully distinguished by our data due to the large uncertainties.
    }
    \label{fig:N2}
\end{figure}

Because our 2D fits naturally yield $e_{10}$, the continuum $N^{-2}$ contribution to $E/N^2$ at each temperature, we can extract the NLO $N$ dependence in \eqref{internal_energy}.
In other words, getting values of $e_{10}$ allows us to fit $E_1(T)$ in \eqref{internal_energy}.

We attempted four fits of the parameters $b_0$ (known exactly from string theory, and approximately $-5.77$) and $b_1$: a fit of $b_0$ only, a fit of $b_1$ fixing $b_0$ to its known value, and a simultaneous fit of both parameters, as well as a fit to a generic power law $b_0 T^p$.  
Because our data point at $T=0.4$ has not been determined with high accuracy, we do not include it in the fits.
All fit forms do a good job describing our data, due to the large uncertainties in the values of $e_{10}$, and can be seen in \figref{N2}. 

We observe general consistency with the known values, but cannot confidently extract precision values.
The two-parameter fit yields $b_0=-5.8\pm3.0$---the central value reproducing the known value, but with $\sim50\%$ uncertainty---and a very large uncertainty on $b_1=-3.4\pm5.7$.
However, the central value of $b_1$ for that fit is concordant with central value of the fit with fixed $b_0$, $-3.5\pm2.0$, which may give a modicum of confidence to the two-parameter fit.

Smaller uncertainties on the data points and lower temperatures are required to perform precision comparisons between the gravity and gauge theory at \order{N^{-2}}.
One strategy might be to calculate at smaller $N$ to enhance the correction terms, but, unfortunately, it is difficult to probe small $N$ because the Monte Carlo simulations find flat directions more quickly.
This clearly hinders our ability to observe effects beyond \order{N^{-2}}.\footnote{Note however that a method introduced and used in \Ref{Hanada:2013rga} might help in the current parameter region.}

\section{Discussion}

We have started a systematic, large-scale lattice simulation of the D0-brane quantum mechanics.
In particular, we have performed the extrapolation of the internal energy to the continuum limit and to large $N$ in a wide range of temperatures.
This enabled us to do a precision test of the gauge/gravity duality. 
By assuming the form of the temperature dependence coming from supergravity calculations
\begin{align}
\label{eq:en2again}
    \frac{E}{N^2}
    = 
        \frac{\left(    a_0 T^{14/5} + a_1 T^{23/5} + a_2 T^{29/5} + a_3T^{32/5}\cdots    \right)}{N^0}
    +   \frac{\left(    b_0 T^{2/5} + b_1 T^{11/5} +               \cdots    \right)}{N^2}
    +   \order{\frac{1}{N^4}}
\end{align}
we can check the agreement between supergravity and the D0-brane quantum mechanics, where results for the latter come from lattice Monte Carlo simulations.
The fit results are summarized in \tabref{results}. 
 
\begin{table}[h]
    \begin{center}
    \begin{tabular}{>{$}c<{$}r@{$\pm$}lr@{$\pm$}l||c>{$}c<{$}r@{$\pm$}lr@{$\pm$}l}
        & \multicolumn{2}{c}{free} & \multicolumn{2}{c||}{$a_0$ fixed} & & 
        & \multicolumn{2}{c}{free} & \multicolumn{2}{c}{$b_0$ fixed} \\\hline
        a_0 & 7.4 & 0.5 & \multicolumn{2}{c||}{7.41}    & & b_0 & $-$5.8 & 3.0 & \multicolumn{2}{c}{$-$5.77}          \\
        a_1 & $-$9.7 & 2.2 & $-$10.0 & 0.4                  & & b_1 & $-$3.4  & 5.7 & $-$3.5 & 2.0                          \\
        a_2 & 5.6 & 1.8 & 5.8  & 0.5                  & & 
    \end{tabular}
    \end{center}
    \caption{
    A summary of our fit results for $E/N^2$ parametrized as in \Eqref{en2again}, using only knowledge of the powers of the temperature dependence.
    These results are described in greater detail in Sections~\ref{sec:confirmation},~\ref{sec:nnlo},~and~\ref{sec:N2_physics}.
    The two columns for each quantity are the results from the totally free fit and the fit where the leading behavior is fixed to the known value.
    }
    \label{tab:results}
\end{table}
For the first time we have directly determined the leading coefficient $a_0$ and obtained $a_0=7.4\pm 0.5$, which nicely agrees with 7.41, the value known from the supergravity calculation.
We also determined the next-to-leading temperature dependence $a_1=-10.2\pm2.4$ and $p_1=4.6\pm0.3$ in the continuum limit, and found agreement with \Ref{Kadoh:2015mka}.

The precision of our large-$N$, continuum extrapolated points makes it hard to believe that dramatic improvements can be achieved through larger statistical sampling.
Instead, to reduce our $\sim7\%$ error on $a_0$ we would require more data points or simulations at lower temperatures.
Unfortunately, stabilizing the Monte Carlo simulations at lower temperatures requires going to even larger values of $N$ which is numerically costly.
For the same reason it is challenging to obtain a precise determination of $E_1(T)$, the $1/N^2$ corrections, in the parameter region we considered, where we had to use $N \geq 16$.
However, we were able to extract general agreement with the known $b_0=-5.77$, albeit with sizable uncertainty.

We believe that the current results demonstrate the power of large-scale supercomputer simulations applied to superstring theory.
A number of future directions are definitely worth investigating.
Besides increasing the precision of the numerical results presented in our study to test the duality between the D0-brane quantum mechanics and type IIA superstring even more accurately, we will focus on the very low temperature region where the system is expected to be described by M-theory~\cite{Banks:1996vh,deWit:1988ig,Itzhaki:1998dd}.
Studying super Yang-Mills in other spacetime dimensions with an equally large-scale study is an another important direction.

While testing the duality is crucial, framing quantum gravitational puzzles in terms of the gauge theory could be especially rewarding.
Can one see the emergence of the bulk spacetime from the gauge theory?
Is there a firewall?
Is it possible to explicitly trace the unitary evaporation of the black hole?
Fascinating frontiers lie ahead.

\section*{Acknowledgements}
\addcontentsline{toc}{section}{Acknowledgements}
M.~H. would like to thank Y.~Hyakutake for discussions.
The work of M.~H. is supported in part by the Grant-in-Aid of the Japanese Ministry of Education, Sciences and Technology, Sports and Culture (MEXT) for Scientific Research (No.~25287046).
The work of G.~I. was supported, in part, by Program to Disseminate Tenure Tracking System, MEXT, Japan and by KAKENHI~(16K17679).
S.~S. was supported by the MEXT-Supported Program for the Strategic Research Foundation at Private Universities ``Topological Science'' (Grant No.~S1511006)
This work was performed under the auspices of the U.S. Department of Energy by Lawrence Livermore National Laboratory under contract~{DE-AC52-07NA27344}.
Numerical calculations were performed on the Vulcan BlueGene/Q at LLNL, supported by the LLNL Multiprogrammatic and Institutional Computing program through a Tier 1 Grand Challenge award and on the RIKEN K Computer.

\bibliographystyle{unsrt}
\addcontentsline{toc}{section}{References}
\bibliography{bfss}

\newpage
\appendix
\section{Correlations Between Polyakov Loop and Internal Energy}\label{sec:correlations}
In this section we analyze the possible presence of correlations in the distributions of the Polyakov loop $|P|$ (cfr. \eqref{polyloopabs}) and of the internal energy $E/N^2$ (cfr. \eqref{internalenergy}).
Our aim is to test how similar the distributions of $|P|$ are for different values of $E/N^2$.
To compare distributions of samples drawn in the same Monte Carlo simulation, it is useful to apply the two-sample Kolmogorov-Smirnov (KS) test.
This test is designed to give a statistical measure to the similarity of two distributions.

In practice, given two data sets of size $n_1$ and $n_2$ with their corresponding empirical distribution functions $F^{(1)}_{n_1}(x)$ and $F^{(2)}_{n_2}(x)$, the KS statistic is
\begin{equation}
  \label{eq:kstest}
  D(n_1,n_2) \equiv D \; = \; \textrm{sup}_x |F^{(1)}_{n_1}(x) - F^{(2)}_{n_2}(x)| \ .
\end{equation}
This statistic tests the hypothesis that the distributions of the two data sets are the same.
If $D(n_1,n_2)$ is larger than a critical value, the distributions are not the same with an associated confidence level.
Often the confidence level is expressed in terms of a p-value.\footnote{The p-value is a well-known tool used in frequentist statistics. 
It is the probability of finding the observed, or more extreme, results when the hypothesis is true.
Large p-values support the hypothesis.}
We use the KS test and reject the hypothesis that two $|P|$ distributions are the same if the associated p-value is less than 1\%.

We compare two distributions that have the same number of samples--- for example $1/12 \sim 8\%$ of the whole set.
Therefore, we start from our Monte Carlo history of $E/N^2$ and select 8\% of the configurations with the smallest energy.
Then we select the next 8\% and so on.
For each 8\% bin of configurations, we look at the distribution of the Polyakov loop and compare them pair-wise.
We also repeat this analysis with bins containing only 4\% of the configurations each, noticing no significant differences.

\begin{figure}[hp]
  \centering
    \includegraphics[width=.5\textwidth]{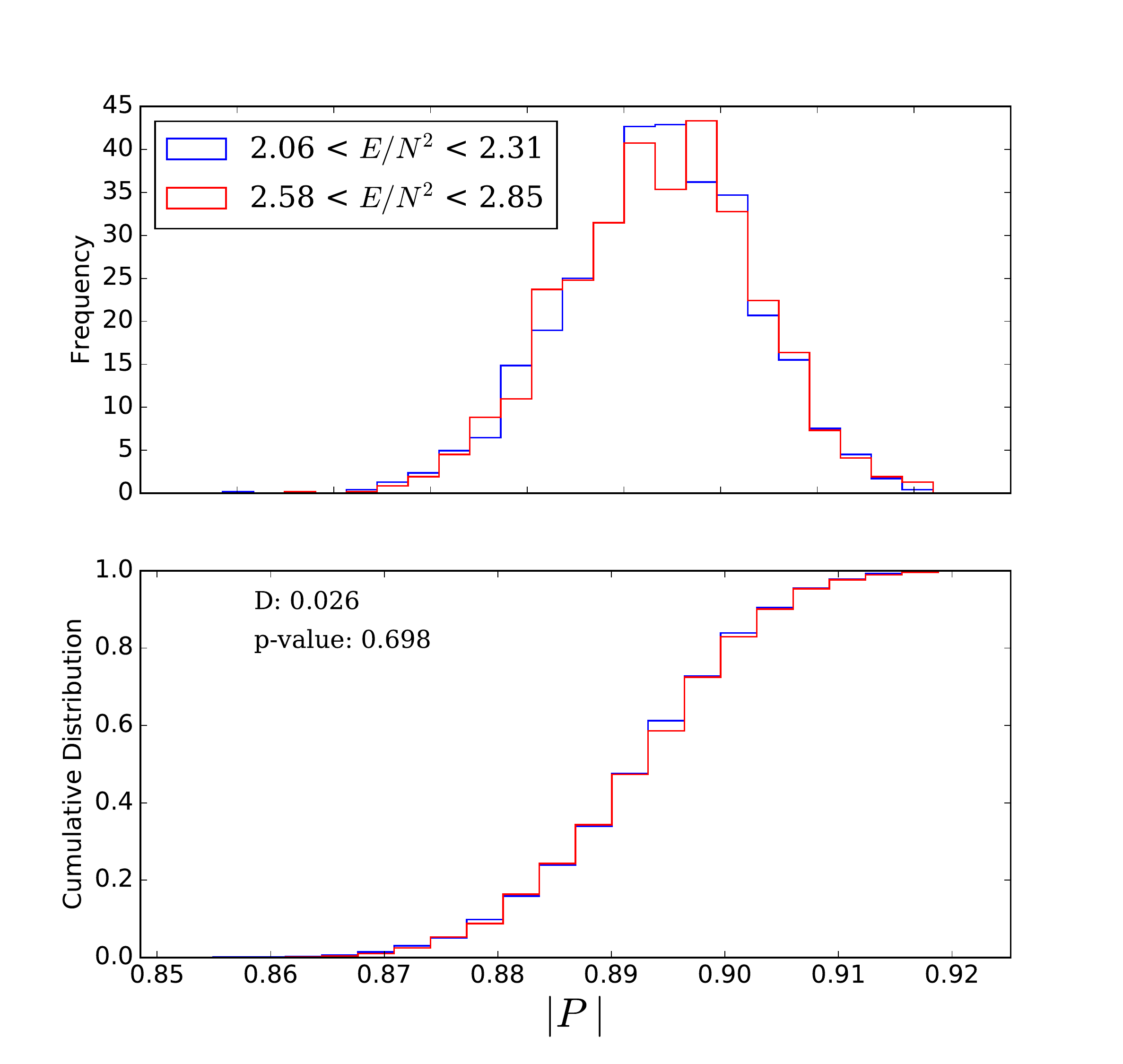}
  \caption{ The distributions of $|P|$ at $N=16$, $L=16$ and $T=0.8$ for different energy intervals reported in the legend.
    These energy intervals are close to the average value of $E/N^2$ for this ensemble.
    The upper panel shows the normalized probability distribution, while the lower panel shows the cumulative distribution as a proxy for the empirical distribution function used in the KS test.
    The value of the KS statistic $D$ and its associated p-value are also shown, giving confidence that the two underlying distributions are the same.
    }\label{fig:KStest1}
\end{figure}

\begin{figure}[hp]
  \centering
    \includegraphics[width=.5\textwidth]{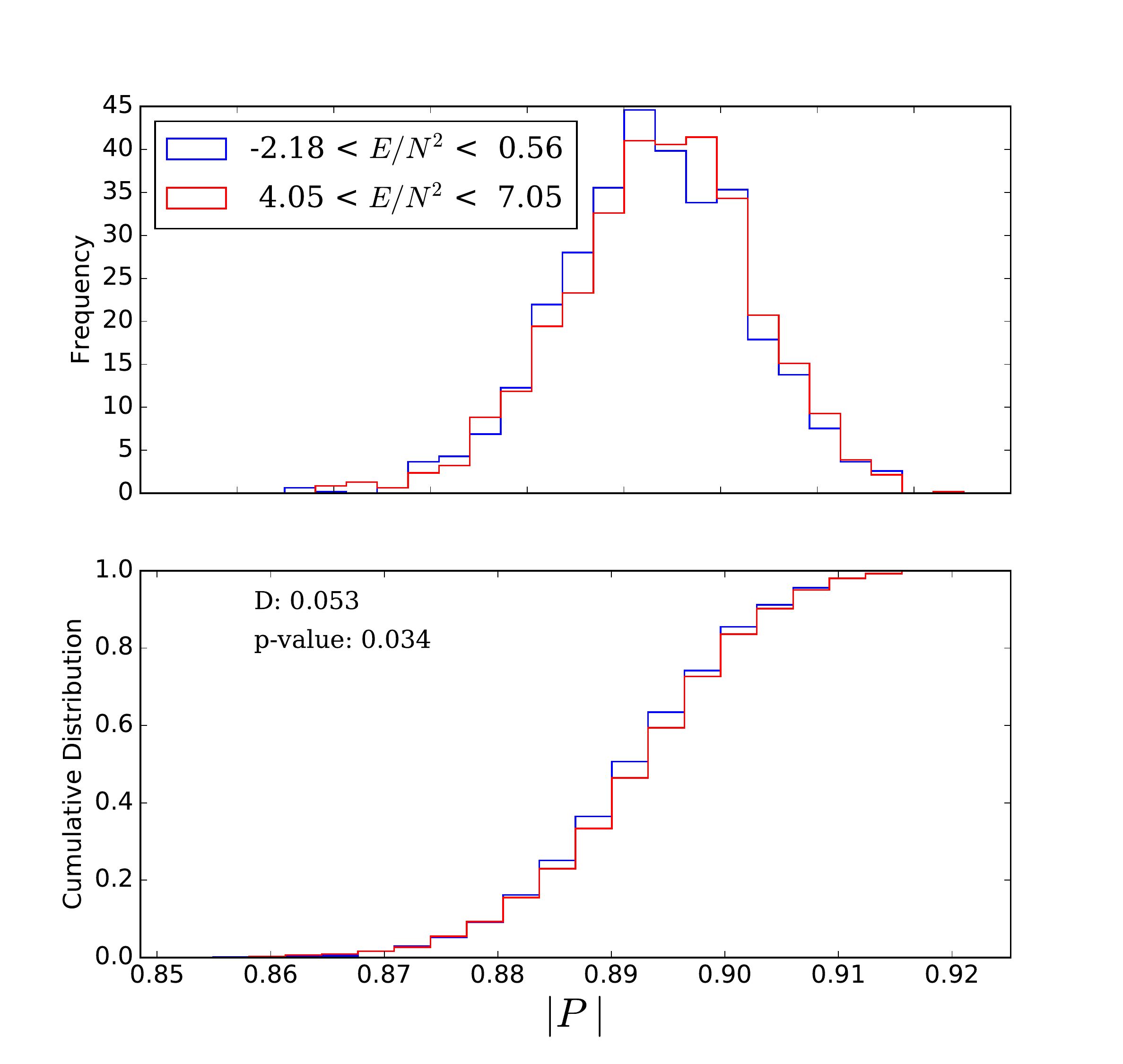}
  \caption{ Same as \Figref{KStest1}, but for energy intervals at the tail of the energy distribution, instead of around the average.
    The KS statistic is larger and the p-value is considerably lower.
    }\label{fig:KStest2}
\end{figure}

For example, in the case of our $N=16$, $L=16$ and $T=0.8$ ensemble we find out that the test is successful $>92\%$ of the time.
Two distributions passing the test are shown in \figref{KStest1}.
We also notice that the failing tests occur for energy intervals near the tail of the energy distribution.
In such cases it is clearly harder to obtain a faithful sampling of the distribution.
Moreover, an equivalence of the Polyakov loop distributions in the tail of the energy fluctuations is not strictly required to corroborate our argument in \Secref{phasequench} about the validity of the phase-quenched approximation of the Pfaffian.
An example of the KS test in the tail of the energy distribution is shown in \figref{KStest2}.

\afterpage{%
  \begin{landscape}
\section{Lattice Measurements}\label{sec:measurements}
Here we present a summary of our ensembles and corresponding measured observables.
    \begin{center}
\begin{longtable}{
    >{$}r<{$}
    >{$}r<{$}
    >{$}r<{$}
    r
    r
    >{$}r<{$}@{$\pm$}>{$}l<{$}
    >{$}r<{$}@{$\pm$}>{$}l<{$}
    >{$}r<{$}@{$\pm$}>{$}l<{$}
    >{$}r<{$}@{$\pm$}>{$}l<{$}
    }
\hline
     T &  N &  L &     action & \Ncfg  &     \multicolumn{2}{c}{$E/N^2$} &  \multicolumn{2}{c}{$|P|$} & \multicolumn{2}{c}{$R^2$}   & \multicolumn{2}{c}{$F^2$}\\
\hline
\hline
\endhead
\hline
\endfoot
   0.4 & 24 & 16 &   improved & 15935 & 0.827 & 0.005 & 0.72770 & 0.00035 & 3.2504 & 0.0015 & 14.530 & 0.002 \\
       &    & 24 &   improved &  2321 & 0.719 & 0.031 & 0.72888 & 0.00129 & 3.3459 & 0.0039 & 15.627 & 0.011 \\
       &    & 32 &   improved &  6625 & 0.657 & 0.027 & 0.72721 & 0.00116 & 3.4110 & 0.0020 & 16.319 & 0.008 \\
       & 32 &  8 &   improved &  3057 & 0.903 & 0.009 & 0.74150 & 0.00348 & 4.8789 & 0.0967 & 13.846 & 0.006 \\
       &    & 12 &   improved &  2491 & 0.907 & 0.010 & 0.72754 & 0.00089 & 3.1663 & 0.0024 & 13.651 & 0.005 \\
       &    & 16 &   improved &  8242 & 0.835 & 0.007 & 0.72732 & 0.00054 & 3.2387 & 0.0012 & 14.518 & 0.003 \\
       &    & 24 &   improved &  1331 & 0.692 & 0.052 & 0.72919 & 0.00453 & 3.3414 & 0.0025 & 15.635 & 0.012 \\
       &    & 32 &   improved &  1888 & 0.629 & 0.029 & 0.72849 & 0.00142 & 3.4016 & 0.0018 & 16.311 & 0.008 \\ \hline
   0.5 & 16 &  8 &   improved & 21101 & 1.229 & 0.004 & 0.78847 & 0.00031 & 3.1104 & 0.0026 & 13.068 & 0.003 \\
       &    & 12 &   improved & 17201 & 1.140 & 0.007 & 0.79566 & 0.00032 & 3.2304 & 0.0014 & 14.374 & 0.003 \\
       &    & 16 &   improved & 17933 & 1.081 & 0.009 & 0.79599 & 0.00035 & 3.3086 & 0.0012 & 15.207 & 0.004 \\
       &    & 32 &   improved & 15101 & 0.907 & 0.020 & 0.79689 & 0.00049 & 3.4747 & 0.0017 & 16.897 & 0.006 \\
       & 24 &  8 &   improved & 20951 & 1.243 & 0.004 & 0.78964 & 0.00028 & 3.0776 & 0.0007 & 13.038 & 0.002 \\
       &    & 16 &   improved & 19765 & 1.092 & 0.006 & 0.79718 & 0.00020 & 3.2883 & 0.0005 & 15.194 & 0.002 \\
       &    & 24 &   improved & 14957 & 0.979 & 0.010 & 0.79741 & 0.00029 & 3.3898 & 0.0006 & 16.240 & 0.003 \\
       &    & 32 &   improved & 10469 & 0.941 & 0.024 & 0.79727 & 0.00051 & 3.4457 & 0.0012 & 16.851 & 0.007 \\
       & 32 &  8 &   improved & 16253 & 1.248 & 0.003 & 0.78995 & 0.00020 & 3.0712 & 0.0006 & 13.032 & 0.002 \\
       &    & 12 &   improved &  3569 & 1.155 & 0.010 & 0.79600 & 0.00049 & 3.2012 & 0.0010 & 14.357 & 0.004 \\
       &    & 16 &   improved &  7885 & 1.093 & 0.009 & 0.79730 & 0.00034 & 3.2830 & 0.0007 & 15.196 & 0.003 \\
       &    & 24 &   improved &  2873 & 0.946 & 0.047 & 0.79852 & 0.00123 & 3.3815 & 0.0032 & 16.223 & 0.012 \\
       &    & 32 &   improved &  5469 & 0.955 & 0.023 & 0.79833 & 0.00044 & 3.4386 & 0.0011 & 16.841 & 0.006 \\ \hline
   0.6 & 16 &  8 &   improved & 27221 & 1.560 & 0.005 & 0.83423 & 0.00018 & 3.1410 & 0.0006 & 13.728 & 0.002 \\
       &    & 12 &   improved & 19051 & 1.475 & 0.007 & 0.84077 & 0.00021 & 3.2708 & 0.0008 & 15.001 & 0.003 \\
       &    & 16 &   improved & 18141 & 1.432 & 0.010 & 0.84156 & 0.00023 & 3.3477 & 0.0010 & 15.790 & 0.004 \\
       &    & 24 &   improved &  8977 & 1.339 & 0.021 & 0.84184 & 0.00034 & 3.4410 & 0.0020 & 16.754 & 0.008 \\
   0.6 & 16 & 32 &   improved & 18677 & 1.267 & 0.021 & 0.84181 & 0.00028 & 3.4951 & 0.0014 & 17.327 & 0.006 \\
       & 24 &  8 &   improved & 23971 & 1.569 & 0.004 & 0.83474 & 0.00017 & 3.1290 & 0.0005 & 13.731 & 0.002 \\
       &    & 12 &   improved & 19171 & 1.481 & 0.007 & 0.84083 & 0.00018 & 3.2602 & 0.0007 & 15.012 & 0.003 \\
       &    & 16 &   improved & 19961 & 1.429 & 0.008 & 0.84205 & 0.00018 & 3.3349 & 0.0006 & 15.790 & 0.003 \\
       &    & 24 &   improved & 25249 & 1.346 & 0.009 & 0.84176 & 0.00015 & 3.4262 & 0.0006 & 16.753 & 0.003 \\
       &    & 32 &   improved & 12577 & 1.276 & 0.025 & 0.84212 & 0.00030 & 3.4780 & 0.0012 & 17.309 & 0.007 \\
       & 32 &  8 &   improved & 19017 & 1.575 & 0.005 & 0.83539 & 0.00024 & 3.1248 & 0.0007 & 13.731 & 0.003 \\
       &    & 16 &   improved & 10071 & 1.442 & 0.009 & 0.84182 & 0.00022 & 3.3306 & 0.0006 & 15.787 & 0.004 \\ \hline
   0.7 & 16 &  8 &   improved & 30641 & 1.959 & 0.005 & 0.86564 & 0.00013 & 3.1941 & 0.0006 & 14.377 & 0.003 \\
       &    & 12 &   improved & 20051 & 1.885 & 0.008 & 0.87096 & 0.00014 & 3.3145 & 0.0008 & 15.579 & 0.004 \\
       &    & 16 &   improved & 20187 & 1.843 & 0.011 & 0.87181 & 0.00015 & 3.3891 & 0.0009 & 16.333 & 0.005 \\
       &    & 24 &   improved & 10605 & 1.763 & 0.022 & 0.87126 & 0.00024 & 3.4702 & 0.0018 & 17.214 & 0.009 \\
       &    & 32 & unimproved & 21633 & 2.344 & 0.031 & 0.86981 & 0.00017 & 3.5681 & 0.0015 & 18.615 & 0.007 \\
       &    &    &   improved & 19921 & 1.672 & 0.023 & 0.87191 & 0.00021 & 3.5193 & 0.0014 & 17.739 & 0.007 \\
       & 24 &  8 &   improved & 20701 & 1.968 & 0.005 & 0.86574 & 0.00014 & 3.1854 & 0.0005 & 14.385 & 0.003 \\
       &    & 12 &   improved & 19997 & 1.893 & 0.008 & 0.87095 & 0.00013 & 3.3088 & 0.0007 & 15.601 & 0.004 \\
       &    & 16 &   improved & 21451 & 1.849 & 0.007 & 0.87203 & 0.00012 & 3.3789 & 0.0006 & 16.338 & 0.004 \\
       &    & 24 &   improved & 28925 & 1.755 & 0.011 & 0.87126 & 0.00012 & 3.4634 & 0.0007 & 17.237 & 0.004 \\
       &    & 32 &   improved & 16135 & 1.682 & 0.025 & 0.87186 & 0.00019 & 3.5069 & 0.0012 & 17.726 & 0.008 \\
       & 32 &  8 &   improved & 18989 & 1.966 & 0.006 & 0.86612 & 0.00017 & 3.1837 & 0.0007 & 14.394 & 0.004 \\
       &    & 16 &   improved & 10849 & 1.850 & 0.010 & 0.87187 & 0.00015 & 3.3771 & 0.0008 & 16.341 & 0.004 \\ \hline
   0.8 & 16 &  8 & unimproved & 19281 & 3.674 & 0.009 & 0.89048 & 0.00013 & 3.5758 & 0.0028 & 17.845 & 0.008 \\
       &    &    &   improved & 19171 & 2.400 & 0.008 & 0.88790 & 0.00012 & 3.2468 & 0.0007 & 14.994 & 0.004 \\
       &    & 12 &   improved & 22001 & 2.356 & 0.009 & 0.89220 & 0.00011 & 3.3601 & 0.0008 & 16.138 & 0.005 \\
       &    & 16 & unimproved & 24081 & 3.283 & 0.016 & 0.89113 & 0.00011 & 3.5305 & 0.0011 & 18.305 & 0.006 \\
       &    &    &   improved & 21421 & 2.309 & 0.012 & 0.89318 & 0.00011 & 3.4295 & 0.0010 & 16.856 & 0.005 \\
       &    & 24 & unimproved & 21591 & 3.000 & 0.025 & 0.89069 & 0.00012 & 3.5588 & 0.0012 & 18.663 & 0.007 \\
       &    &    &   improved & 17521 & 2.214 & 0.019 & 0.89292 & 0.00014 & 3.5070 & 0.0013 & 17.687 & 0.008 \\
       &    & 32 & unimproved & 14157 & 2.710 & 0.044 & 0.89119 & 0.00017 & 3.5803 & 0.0016 & 18.880 & 0.010 \\
   0.8 & 16 & 32 &   improved & 20187 & 2.107 & 0.024 & 0.89261 & 0.00015 & 3.5452 & 0.0014 & 18.131 & 0.009 \\
       & 24 &  8 &   improved & 22151 & 2.417 & 0.006 & 0.88779 & 0.00011 & 3.2416 & 0.0006 & 15.010 & 0.003 \\
       &    & 12 &   improved & 18175 & 2.346 & 0.009 & 0.89207 & 0.00011 & 3.3553 & 0.0008 & 16.155 & 0.005 \\
       &    & 16 &   improved & 20721 & 2.303 & 0.010 & 0.89317 & 0.00009 & 3.4232 & 0.0007 & 16.868 & 0.004 \\
       &    & 24 &   improved &  9153 & 2.216 & 0.020 & 0.89239 & 0.00015 & 3.4968 & 0.0012 & 17.681 & 0.008 \\
       &    & 32 &   improved & 13867 & 2.176 & 0.028 & 0.89287 & 0.00015 & 3.5366 & 0.0015 & 18.134 & 0.009 \\
       & 32 &  8 &   improved & 19213 & 2.424 & 0.007 & 0.88809 & 0.00014 & 3.2394 & 0.0008 & 15.012 & 0.004 \\
       &    & 16 &   improved & 13495 & 2.340 & 0.017 & 0.89302 & 0.00018 & 3.4216 & 0.0014 & 16.869 & 0.008 \\ \hline
   0.9 & 16 &  8 & unimproved & 20411 & 4.221 & 0.010 & 0.90519 & 0.00009 & 3.5113 & 0.0012 & 17.926 & 0.005 \\
       &    &    &   improved & 20401 & 2.895 & 0.009 & 0.90433 & 0.00009 & 3.3002 & 0.0008 & 15.599 & 0.004 \\
       &    & 12 &   improved & 20501 & 2.863 & 0.011 & 0.90783 & 0.00009 & 3.4048 & 0.0009 & 16.679 & 0.005 \\
       &    & 16 & unimproved & 37701 & 3.710 & 0.014 & 0.90689 & 0.00007 & 3.5475 & 0.0007 & 18.635 & 0.005 \\
       &    &    &   improved & 21461 & 2.796 & 0.014 & 0.90856 & 0.00009 & 3.4717 & 0.0010 & 17.377 & 0.006 \\
       &    & 24 & unimproved & 21851 & 3.450 & 0.028 & 0.90639 & 0.00010 & 3.5807 & 0.0013 & 19.001 & 0.008 \\
       &    &    &   improved & 17955 & 2.716 & 0.021 & 0.90826 & 0.00011 & 3.5400 & 0.0014 & 18.139 & 0.009 \\
       &    & 32 & unimproved & 12221 & 3.243 & 0.049 & 0.90720 & 0.00013 & 3.6037 & 0.0019 & 19.209 & 0.012 \\
       &    &    &   improved & 14601 & 2.673 & 0.031 & 0.90849 & 0.00013 & 3.5750 & 0.0017 & 18.545 & 0.011 \\
       & 24 &  8 &   improved & 24331 & 2.931 & 0.006 & 0.90413 & 0.00008 & 3.2958 & 0.0006 & 15.613 & 0.004 \\
       &    & 12 &   improved & 17557 & 2.872 & 0.010 & 0.90785 & 0.00009 & 3.4025 & 0.0009 & 16.707 & 0.006 \\
       &    & 16 &   improved & 24441 & 2.822 & 0.010 & 0.90861 & 0.00007 & 3.4658 & 0.0007 & 17.384 & 0.005 \\
       &    & 24 &   improved &  8917 & 2.761 & 0.023 & 0.90774 & 0.00013 & 3.5349 & 0.0014 & 18.156 & 0.009 \\
       &    & 32 &   improved & 16709 & 2.719 & 0.028 & 0.90824 & 0.00010 & 3.5663 & 0.0014 & 18.538 & 0.009 \\
       & 32 &  8 &   improved & 18695 & 2.931 & 0.008 & 0.90441 & 0.00011 & 3.2948 & 0.0009 & 15.620 & 0.005 \\
       &    & 16 &   improved & 12061 & 2.835 & 0.019 & 0.90836 & 0.00016 & 3.4643 & 0.0016 & 17.388 & 0.009 \\ \hline
   1.0 & 16 &  8 & unimproved & 21291 & 4.719 & 0.011 & 0.91705 & 0.00007 & 3.5139 & 0.0010 & 18.245 & 0.006 \\
       &    &    &   improved & 20641 & 3.439 & 0.010 & 0.91672 & 0.00008 & 3.3515 & 0.0009 & 16.185 & 0.005 \\
       &    & 12 &   improved & 20751 & 3.397 & 0.012 & 0.91968 & 0.00008 & 3.4485 & 0.0010 & 17.217 & 0.006 \\
       &    &    & unimproved & 25379 & 4.185 & 0.019 & 0.91907 & 0.00007 & 3.5769 & 0.0010 & 19.034 & 0.007 \\
       &    & 16 &   improved & 21641 & 3.378 & 0.015 & 0.92033 & 0.00008 & 3.5115 & 0.0011 & 17.876 & 0.007 \\
   1.0 & 16 & 24 & unimproved & 23391 & 3.937 & 0.030 & 0.91855 & 0.00008 & 3.6099 & 0.0013 & 19.376 & 0.009 \\
       &    &    &   improved & 17469 & 3.290 & 0.024 & 0.91979 & 0.00009 & 3.5755 & 0.0017 & 18.600 & 0.011 \\
       &    & 32 & unimproved & 13503 & 3.690 & 0.054 & 0.91895 & 0.00012 & 3.6323 & 0.0019 & 19.582 & 0.013 \\
       &    &    &   improved & 15555 & 3.185 & 0.034 & 0.91985 & 0.00011 & 3.6068 & 0.0018 & 18.971 & 0.013 \\
       &    & 48 & unimproved & 12026 & 3.534 & 0.092 & 0.92064 & 0.00014 & 3.6592 & 0.0028 & 19.810 & 0.021 \\
       &    &    &   improved &  5772 & 3.154 & 0.094 & 0.92101 & 0.00025 & 3.6431 & 0.0042 & 19.401 & 0.027 \\
       &    & 64 & unimproved & 15024 & 3.505 & 0.112 & 0.92051 & 0.00015 & 3.6808 & 0.0029 & 19.999 & 0.020 \\
       &    &    &   improved &  7280 & 3.210 & 0.105 & 0.92070 & 0.00021 & 3.6752 & 0.0042 & 19.715 & 0.030 \\
       & 24 &  8 &   improved & 22605 & 3.467 & 0.007 & 0.91663 & 0.00007 & 3.3490 & 0.0007 & 16.208 & 0.004 \\
       &    & 12 &   improved & 19817 & 3.413 & 0.011 & 0.91956 & 0.00007 & 3.4478 & 0.0009 & 17.251 & 0.006 \\
       &    & 16 &   improved & 20655 & 3.363 & 0.013 & 0.92041 & 0.00007 & 3.5084 & 0.0009 & 17.901 & 0.006 \\
       &    & 24 &   improved & 21725 & 3.283 & 0.019 & 0.91964 & 0.00008 & 3.5705 & 0.0011 & 18.609 & 0.008 \\
       &    & 32 &   improved & 19383 & 3.268 & 0.028 & 0.91997 & 0.00009 & 3.5993 & 0.0014 & 18.964 & 0.010 \\
       & 32 &  8 &   improved & 17881 & 3.469 & 0.008 & 0.91684 & 0.00009 & 3.3482 & 0.0009 & 16.216 & 0.006 \\
       &    & 16 &   improved & 14915 & 3.386 & 0.019 & 0.92018 & 0.00011 & 3.5078 & 0.0013 & 17.902 & 0.009 \\
\end{longtable}

    \end{center}
  \end{landscape}
}

\afterpage{%
  \begin{landscape}
\section{Simultaneous Continuum Large-$N$ Extrapolations}\label{sec:extrap_full}

Here we give a more complete version \Tabref{extrap}, summarizing our simultaneous extrapolation to the continuum, large-$N$ limit via \eqref{2D_fit_form}, including the lattice-spacing effects and all the off-diagonal entries of the covariance matrix.

    \begin{center}
\begin{tabular}{
    c
    >{$}r<{$}@{$\pm$}>{$}l<{$}
    >{$}r<{$}@{$\pm$}>{$}l<{$}
    >{$}r<{$}@{$\pm$}>{$}l<{$}
    >{$}r<{$}@{$\pm$}>{$}l<{$}
    >{$}c<{$}
    >{$}c<{$}
    >{$}c<{$}
    >{$}c<{$}
    >{$}c<{$}
    >{$}c<{$}
    rc}
    \hline
    $T$     &   \multicolumn{2}{c}{$e_{00}$}    &   \multicolumn{2}{c}{$-e_{10}$}    & \multicolumn{2}{c}{$e_{01}$}    & \multicolumn{2}{c}{$-e_{02}$} & \Sigma_{00,10} & \Sigma_{00,01} & \Sigma_{00,02}   & \Sigma_{10,01} & \Sigma_{10,02} & \Sigma_{01,02} & $\chi^2$      & DOF    \\\hline\hline
0.4 & 0.38 & 0.06 &  5.4 & 9.2 & 10.0 & 1.8 & 44 & 15 & -0.1631 & -0.11 & 0.79  & -0.36 & 15.9 & -28 &  1.3 &  4\\
0.5 & 0.74 & 0.02 &  6.7 & 1.5 &  7.2 & 0.6 & 25 &  3 & +0.0005 & -0.01 & 0.08  & -0.14 & \phantom{1}0.8 &  -2  &  7.2 &  9\\
0.6 & 1.15 & 0.02 &  5.0 & 1.8 &  5.8 & 0.6 & 19 &  3 & -0.0046 & -0.01 & 0.08  & -0.07 & \phantom{1}0.4 &  -2  &  8.8 &  8\\
0.7 & 1.54 & 0.03 &  3.9 & 2.0 &  6.4 & 0.7 & 23 &  3 & -0.0066 & -0.02 & 0.09  & -0.07 & \phantom{1}0.4 &  -2  &  8.8 &  8\\
0.8 & 1.99 & 0.03 &  6.2 & 2.5 &  7.0 & 0.8 & 28 &  4 & -0.0151 & -0.02 & 0.13  & -0.09 & \phantom{1}0.8 &  -3  & 15.1 &  8\\
0.9 & 2.57 & 0.04 & 11.9 & 2.9 &  5.9 & 0.9 & 23 &  4 & -0.0192 & -0.03 & 0.17  & -0.08 & \phantom{1}0.8 &  -4  &  3.3 &  8\\
1.0 & 3.11 & 0.04 &  8.4 & 3.2 &  5.9 & 0.9 & 23 &  4 & -0.0218 & -0.03 & 0.16  & -0.15 & \phantom{1}1.3 &  -4  &  8.9 & 10\\\hline
\end{tabular}

    \end{center}

    If measurements of variables $x_i$ are normally distributed, then na\"ively their joint probability distribution $P$ might be
    \begin{equation}
        P(x) = c\; e^{-\frac{1}{2}\sum_i \left(\frac{x_i-\mu_i}{\sigma_i}\right)^2}
    \end{equation}
    where $\mu_i$ represents the central value and $\sigma_i$ the spread in the measurement of (ie. the uncertainty of) variable $x_i$, and $c$ is a normalization constant.
    However, more generically the measurements might be distributed according to
    \begin{equation}
        P(x) = c\; e^{-\frac{1}{2}(x-\mu)_i\; \Sigma^{-1}_{ij}\; (x-\mu)_j}
    \end{equation}
    where $\Sigma$ is the \emph{covariance matrix}, a symmetric positive-definite matrix.
    The uncertainty on a single variable $x_i$ is $\sqrt{\Sigma_{ii}}$.
    An off-diagonal entry in the covariance matrix indicates how the errors on the two variables $i$ and $j$ are correlated.
    We use the shorthand that the two subscripts on $\Sigma$ are the subscripts on the corresponding $e$ variables.
    
    We determine the best fit by minimizing the usual $\chi^2$ fit metric.
    We fix the covariant errors by finding the values of the variables where the minimal $\chi^2$ increases by 1 (or, equivalently, where $P$ decreases by $1/e$).

  \end{landscape}
}

\end{document}